 \documentclass[twocolumn]{aastex631}

\usepackage{graphicx}
\graphicspath{ {./results/} }
\usepackage{hyperref,lipsum}
\usepackage{booktabs} 
\usepackage{amssymb}

\usepackage{multirow}

\usepackage{graphicx}
\newcommand{\rhead}[1]{\rotatebox{90}{\scriptsize #1}}

\renewcommand{\arraystretch}{1.2}

\usepackage{comment} 
\begin{document}

\title{SENSITIVITY OF NUCLEAR REACTION RATES IN X-RAY BURST MODELS}

\author[0000-0002-5476-0369]{I. Sultana}
\affiliation{Department of Physics, Central Michigan University, Mount Pleasant, MI 48859, USA}
\affiliation{Joint Institute for Nuclear Astrophysics-Center for the Evolution of the Elements (JINA-CEE), USA}

\author[0000-0002-0537-9043]{A. Estradé}
\affiliation{Department of Physics, Central Michigan University, Mount Pleasant, MI 48859, USA}
\affiliation{Joint Institute for Nuclear Astrophysics-Center for the Evolution of the Elements (JINA-CEE), USA}

\author[0000-0001-6307-9818]{B. S. Meyer}
\affiliation{Department of Physics and Astronomy, Clemson University, Clemson, SC 29634-0978, USA}

\author[0000-0003-1674-4859]{H. Schatz}
\affiliation{Joint Institute for Nuclear Astrophysics-Center for the Evolution of the Elements (JINA-CEE), USA}
\affiliation{Department of Physics and Astronomy, Michigan State University, East Lansing,
MI 48824, USA}
\affiliation{Facility for Rare Isotope Beams, Michigan State University, East Lansing, Michigan 48824, USA}




\begin{abstract}

Type I X-ray bursts (XRBs) are thermonuclear runaways on the surface of accreting neutron stars, powered by rapid proton-capture and alpha-capture processes on neutron-deficient nuclei. Uncertainties in the corresponding reaction rates remain a major limitation in modeling burst light curves and ashes. We present a systematic study of the sensitivity of XRB models to uncertainties in charged-particle-induced reaction rates across a broad parameter space of accretion rates and fuel compositions in low-mass X-ray binaries. The study proceeds in two stages: ignition conditions are first determined with a semi-analytic framework coupled to a full reaction network, followed by a sensitivity analysis using the ONEZONE model with individual rate variations. We identify 49 reactions that alter the burst light curve and 182 that significantly impact final abundances. Reactions on bottleneck isotopes in the $\alpha p$- and $rp$-process paths strongly affect both observables, while most $(p,\gamma)$ reactions on medium-mass ($A > 32$) and heavy-mass ($A > 55$) nuclei influence only the final composition. Medium-mass cases dominate in He-rich bursts, where the reaction flow terminates earlier, while heavy-mass cases appear in mixed H and He bursts with extended $rp$-process paths reaching $A \approx 110$. We identify a subset of reactions whose rate uncertainties exert influence on the final $^{12}$C yield in helium‐rich bursts, which could have important consequences for the mechanism of ignition of carbon superbursts. Our results identify key targets for nuclear reaction experiments to reduce nuclear physics uncertainties in XRB models.

\end{abstract}

\keywords{}


\section{Introduction} \label{sec:intro}


In low‐mass X‐ray binaries (LMXBs), a neutron star accretes hydrogen‐rich or helium‐rich material from a lower‐mass companion via Roche‐lobe overflow, forming an accretion disk that steadily deposits nuclear fuel onto the stellar surface (\cite{SCHATZ2006601}; \cite{Strohmayer_Bildsten_2006}; \cite{Motta.2011}; \cite{10.1093/mnras/sty757}). As this fresh material is compressed and heated to approximately $10^8$--$10^9$ K under the neutron star's intense gravity, stable burning via the hot CNO cycle can give way to a thermonuclear runaway once cooling can no longer compensate for nuclear energy release (\cite{Fisker_2008}; \cite{2010ApJS..189..204J}). The result is a sudden increase in X-ray luminosity, often exceeding quiescent levels by a factor of ten or more over seconds to minutes, followed by a gradual decay as the synthesized nuclear ashes cool and settle (\cite{1976ApJ...205L.127G}; \cite{1993SSRv...62..223L}; \cite{Strohmayer_Bildsten_2006}; \cite{Chen.2011}; \cite{2021ASSL..461..209G}). These events are known as Type I X-ray bursts (XRBs) and provide a unique astrophysical laboratory for probing dense-matter physics and nuclear reactions in hot and dense environments (\cite{Steiner_2010, 2010ApJS..189..240C, Parikh_2013, Güver_2013, Cyburt_2016, Degenaar2018, Xie_2024}). 



Type I X-ray bursts are powered by a sequence of thermonuclear reactions that synthesize nuclei far from stability and release the energy observed in the burst light curve. The runaway begins with the triple-$\alpha$ process under temperatures of order $0.1$--$0.2$ GK  and quickly transitions into the $\alpha$p-process, where successive $(\alpha,p)$ and $(p,\gamma)$ reactions build nuclei up to mass $A \sim 40$ when the helium fraction remains high (\cite{1981ApJS...45..389W,1994ApJ...432..326V, Woosley_2004, Fisker_2008}). At peak temperatures near $1$--$2$ GK, the rapid proton‐capture (rp) process takes over, driving a fast sequence of $(p,\gamma)$ captures and $\beta^+$ decays along the proton drip line and producing isotopes up to tellurium (\cite{1981ApJS...45..389W, SCHATZ1998167, 2001NuPhA.688..150S, 2010ApJS..189..204J}). The nuclear reaction networks needed to model these processes involve hundreds of unstable nuclei connected by nuclear reactions that, in many cases, have poorly constrained reaction rates. XRB models exhibit strong sensitivities to the uncertainty in these reaction rates, which directly affect both the predicted burst energetics and the composition of the ashes (\cite{Parikh_2008, 2009PhRvC..79d5802P, 2010ApJS..189..240C, Parikh_2013, Cyburt_2016, 2017ApJ...844..139S, Meisel_2019, 10.1093/mnras/staa2484, 2022EPJWC.26011040S}). Hence, improving experimental and theoretical constraints on key reaction rates is essential for accurate models of X-ray bursts and using observed burst properties to probe the physics of neutron stars (\cite{ 2007ApJ...671L.141H, Meisel_2018_1, Lu_2024}). 




%
Although the burst light curve is directly measurable from observations, the makeup of the ashes can only be inferred through models that depend on reliable nuclear physics input.  Recent observations of 4U 1820-30 suggest the possibility of detecting the presence of heavy elements synthesized during the thermonuclear burning of XRBs as absorption and emission features in the bursts' X-ray spectra (\cite{Jaisawal_2025}), which would open tantalizing opportunities to directly  study their nucleosynthetic results.  Of particular interest is the survival of $^{12}$C in the ashes of XRBs, since its abundance not only influences the crustal composition but also determines whether conditions for superburst ignition can be met. Superbursts are rare hours-long thermonuclear flashes that are thought to be triggered by unstable $^{12}$C fusion at column depths of the order of $\sim 10^{12}$ g cm$^{-2}$, requiring a $^{12}$C mass fraction of about 20\% in the ashes (\cite{2001ApJ...559L.127C}, \cite{2006ApJ...646..429C}, \cite{2017symm.conf..121I}). This direct link between the detailed nucleosynthesis of XRBs and the fuel reservoir for superbursts highlights the importance of accurately modeling $^{12}$C composition in the XRB ashes.



Sensitivity studies to explore how variations in nuclear physics data and binary system parameters affect observable outcomes of XRB models have been performed with a variety of techniques, including post-processing, single-zone, and multi-zone simulations. Studies focusing on nuclear physics sensitivity have highlighted the critical need for reducing the nuclear uncertainties in XRB models (\cite{2008NewAR..52..409P, 2009PhRvC..79d5802P, Cyburt_2016, 2017ApJ...844..139S}). The sensitivity studies of  \cite{Cyburt_2016} and \cite{2017ApJ...844..139S} used single-zone models to demonstrate that nuclear physics variations lead to substantial changes in the burst light curves and the composition of the ashes. Building on this work, \cite{Meisel_2018}; \cite{Meisel_2019} investigated how reaction rate uncertainties propagate to neutron star properties, finding that nuclear reaction rate variations can alter crust composition and thermal structure by amounts comparable to those caused by different accretion conditions. While these studies demonstrate the need for accurate nuclear physics data to interpret XRBs' observations through theoretical models, they have focused on a narrow class of LMXBs with material of solar composition accreted at relatively high rates that result in hydrogen-rich XRBs, like the “clocked bursters” GS 1826–238 (\cite{Ubertini_1999}).  

XRBs, however, exhibit a wider range of properties yet to be explored by sensitivity studies. Long-term monitoring by space-based X-ray observatories such as RXTE, INTEGRAL, and NICER has cataloged thousands of XRBs, documenting light curve durations and recurrence intervals, properties that reflect mass-accretion rates spanning from approximately $0.01$ to $\sim0.95\,\dot{m}_{\mathrm{Edd}}$ and fuel mixtures with hydrogen fractions $X_\mathrm{H} \approx 0.1$--$0.7$ and metallicities $Z \approx 0.01$--$0.02$ (\cite{Galloway_2008, 2017PASA...34...19G, 2021ASSL..461..209G, Galloway_2020}). These comprehensive burst catalogs (\cite{Galloway_2020, Alizai.2023}) reveal a diverse range of burst morphologies, inferred accretion and ignition conditions. In parallel, model–observation comparison studies have demonstrated that variations in astrophysical parameters are as important as nuclear physics inputs for reproducing the full diversity of observed burst behavior. For example, \citet{10.1093/mnras/sty757} and \citet{10.1093/mnras/stz2638} showed that matching burst recurrence times, light curve shapes, and energetics requires systematic exploration of source properties such as accreted fuel composition, accretion rate, and neutron star mass. One could expect that nuclear processes and the critical nuclear reactions needed to model these bursts will widely differ, as well. It is therefore crucial to perform sensitivity studies for the full parameter space spanned by X-ray bursting systems to take full advantage of the vast amount of available observational data. This is the goal of the study we present in this article.



Since many of the reactions that drive XRB involve short-lived proton-rich nuclei, direct experimental data are still very limited. A large fraction of the nuclear reaction rates in an XRB reaction network is based on theoretical calculations, leading to uncertainties that in many cases can reach several orders of magnitude (\cite{SCHATZ2006601}, \cite{Parikh_2013}). 
While there has been recent progress with direct measurements of reaction rates (\cite{2020PhRvL.125t2701R, 2023PhRvL.131k2701J}), many critical bottleneck reactions along the $\alpha p$- and rp-process paths remain largely unconstrained. A new generation of radioactive beam facilities has opened opportunities for experiments to significantly reduce the uncertainty of thermonuclear reaction rates for XRB models. Sensitivity studies offer a tool to explore the effect of reaction rate uncertainties over a broad range of isotopes and identify the most influential cases that can be addressed by future experiments.

In this study, we perform a large-scale sensitivity study of Type-I X-ray bursts by combining a semi-analytic ignition model (SETTLE) (\cite{Cumming_2000}), post-processing nucleosynthesis calculations (NucNet Tools) \citep{Meyer:2013iM}, and time-dependent burst simulations (ONEZONE) (\citet{2001NuPhA.688..150S}). We evaluate a grid of 32 binary systems spanning a wide range of accretion rates, hydrogen mass fractions, and metallicities, consistent with values inferred from observations. For each case, we systematically vary 2,744 $(p,\gamma)$, $(\alpha,p)$, and $(\alpha,\gamma)$ reaction rates and quantify their effects on burst light curves and final isotopic abundances. We define two quantitative metrics to assess the impact of variations in reaction rates on burst observables. The modeling framework, nuclear inputs, and sensitivity measures are described in Section~\ref{sec:method}. Results are presented in Section~\ref{sec:results}, and Section~\ref{sec:discussion} discusses the broader trends in rate sensitivities across the 32 binary systems.  

\section{Method} \label{sec:method}

We investigate the effect of nuclear reaction rates on XRB models using an integrated approach that combines a semi-analytic model (SETTLE code) (\cite{Cumming_2000}) with two complementary single-zone reaction network codes, NucNet Tools (\cite{Meyer:2013iM}) and ONEZONE code (\cite{2001NuPhA.688..150S} \cite{Cyburt_2016} \cite{2017ApJ...844..139S}). SETTLE is used to determine the thermodynamic trajectory of the liquid outer layer of a neutron star, and NucNet Tools are used for post-processing calculations to obtain the compositional evolution of the accreted layers. We model the thermonuclear burning during the X-ray bursts and evaluate its sensitivity to nuclear reaction rates using the ONEZONE reaction network code. We quantify the effect of each rate variation using two sensitivity metrics: one that measures deviations in the burst light curve profile and another that assesses shifts in the final nucleosynthetic abundance distribution. It is challenging for the current state-of-the-art full 1D XRB models to perform large-scale sensitivity studies due to computational constraints.  This integrated framework, using single-zone reaction network codes, facilitates computationally efficient sensitivity studies of nuclear reaction rates on XRB models that can cover a broad parameter space of low-mass X-ray binary systems. 
\\

\begin{table*}[ht!]
    \centering
    \begin{tabular}{ccccccccc}
        \toprule
        System & $X_{\mathrm{H,acc}}$ & $Z_{acc}$ & $\dot{m}/\dot{m}_{\mathrm{Edd}}$ & $T_{\mathrm{ign}} (\rm GK)$ & $P_{\mathrm{ign}}\,(\mathrm{erg\,cm^{-3}})$  & $X_{\mathrm{H,ign}}$ & $X_{\mathrm{He,ign}}$ & $Z_{\mathrm {ign}}$\\
        \midrule
        H70Z2M5  & 0.7  & 0.02 & 0.05 & 0.297 & 2.17$\times 10^{22}$ & 0.17  & 0.79 & 0.04 \\
        H70Z2M10 & 0.7  & 0.02 & 0.1  & 0.318 & 2.52$\times 10^{22}$ & 0.39  & 0.57 & 0.04\\
        H70Z2M20 & 0.7  & 0.02 & 0.2  & 0.339 & 2.87$\times 10^{22}$ & 0.52  & 0.44 & 0.04\\
        H70Z2M50 & 0.7  & 0.02 & 0.5  & 0.366 & 3.04$\times 10^{22}$ & 0.62 & 0.34 & 0.04 \\
        H70Z2M90 & 0.7  & 0.02 & 0.9  & 0.39  & 2.98$\times 10^{22}$ & 0.65 & 0.31 & 0.04 \\
        H50Z2M5  & 0.5  & 0.02 & 0.05 & 0.282 & 2.01$\times 10^{22}$ & 0.02 & 0.95 & 0.03\\
        H50Z2M10 & 0.5  & 0.02 & 0.1  & 0.3 & 2.46$\times 10^{22}$ & 0.19 & 0.77 & 0.04\\
        H50Z2M20 & 0.5  & 0.02 & 0.2  & 0.314 & 2.54$\times 10^{22}$ & 0.35  & 0.62 & 0.03 \\
        H50Z2M50 & 0.5  & 0.02 & 0.5  & 0.336 & 2.62$\times 10^{22}$ & 0.43 & 0.53 & 0.04 \\
        H50Z2M90 & 0.5  & 0.02 & 0.9  & 0.357 & 2.62$\times 10^{22}$ & 0.46 & 0.51 & 0.03 \\
        H30Z2M10 & 0.3  & 0.02 & 0.1  & 0.284  & 2.22$\times 10^{22}$ & 0.03 & 0.93 & 0.04\\
        H30Z2M20 & 0.3  & 0.02 & 0.2  & 0.296 & 2.29$\times 10^{22}$ & 0.17 & 0.80 & 0.03\\
        H30Z2M50 & 0.3  & 0.02 & 0.5  & 0.315 & 2.41$\times 10^{22}$ & 0.24 & 0.73 & 0.03\\
        H30Z2M90 & 0.3  & 0.02 & 0.9  & 0.333 & 2.26$\times 10^{22}$ & 0.26 & 0.70 & 0.04  \\
        H10Z2M50 & 0.1  & 0.02 & 0.5  & 0.3 & 2.23$\times 10^{22}$ & 0.05 & 0.92  & 0.03\\
        H10Z2M90 & 0.1  & 0.02 & 0.9  & 0.315 & 2.01$\times 10^{22}$ & 0.07 & 0.90 & 0.03  \\
        H70Z1M5  & 0.7  & 0.01 & 0.05 & 0.29  & 3.05$\times 10^{22}$ & 0.29 & 0.68 & 0.03\\
        H70Z1M10 & 0.7  & 0.01 & 0.1  & 0.308  & 3.49$\times 10^{22}$ & 0.46 & 0.51 & 0.03\\
        H70Z1M20 & 0.7  & 0.01 & 0.2  & 0.327 & 4.03$\times 10^{22}$ & 0.56 & 0.42 & 0.02 \\
        H70Z1M50 & 0.7  & 0.01 & 0.5  & 0.351 & 3.70$\times 10^{22}$  & 0.65 & 0.33 & 0.02 \\
        H70Z1M90 & 0.7  & 0.01 & 0.9  & 0.378 & 3.56$\times 10^{22}$ & 0.67 & 0.31 & 0.02\\
        H50Z1M5  & 0.5  & 0.01 & 0.05 & 0.275 & 2.83$\times 10^{22}$ & 0.14  & 0.84 & 0.02 \\
        H50Z1M10 & 0.5  & 0.01 & 0.1  & 0.288 & 3.14$\times 10^{22}$ & 0.30 & 0.68 & 0.02 \\
        H50Z1M20 & 0.5  & 0.01 & 0.2  & 0.3 & 3.20$\times 10^{22}$ & 0.40 & 0.58 & 0.02\\
        H50Z1M50 & 0.5  & 0.01 & 0.5  & 0.324 & 3.18$\times 10^{22}$ & 0.46 & 0.52 & 0.02 \\
        H50Z1M90 & 0.5  & 0.01 & 0.9  & 0.347 & 2.87$\times 10^{22}$ & 0.48 & 0.50 & 0.02\\
        H30Z1M10 & 0.3  & 0.01 & 0.1  & 0.275  & 2.97$\times 10^{22}$ & 0.12 & 0.86 & 0.02\\
        H30Z1M20 & 0.3  & 0.01 & 0.2  & 0.285 & 3.14$\times 10^{22}$  & 0.20 & 0.78 & 0.02\\
        H30Z1M50 & 0.3  & 0.01 & 0.5  & 0.308 & 2.92$\times 10^{22}$ & 0.26 & 0.72 & 0.02 \\
        H30Z1M90 & 0.3  & 0.01 & 0.9  & 0.326 & 2.45$\times 10^{22}$ & 0.28 & 0.70 & 0.02  \\
        H10Z1M20 & 0.1  & 0.01 & 0.2  & 0.272 & 2.93$\times 10^{22}$ & 0.01 & 0.97 &0.02\\
        H10Z1M50 & 0.1  & 0.01 & 0.5  & 0.294 & 2.81$\times 10^{22}$ & 0.06 & 0.92 & 0.02\\
        \bottomrule
    \end{tabular}
    \caption{Ignition conditions across our 32-system sensitivity grid.Parameters include accreted hydrogen mass fraction ($X_{H,\rm acc}$), accreted metallicity ($Z_{acc}$), and accretion rate (normalized to Eddington, $\dot m / \dot m_{\rm Edd}$), along with the resulting ignition temperature ($T_{\rm ign}$), pressure ($P_{\rm ign}$), and mass fractions of hydrogen ($X_{H,\rm ign}$), helium ($X_{He,\rm ign}$), and metallicity distribution ($Z_{ign}$), after postprocessing with NucNet tools at ignition. All systems adopt a $g_{14} = 1.85$ which corresponds to a neutron star mass of $M = 1.8\,M_\odot$ and radius of $R = 13$~km (with relativistic correction). The system label encodes the parameters as H⟨$X_{H,\rm acc}\times100$⟩Z⟨$Z\times100$⟩M⟨$\dot m / \dot m_{\rm Edd}\times100$⟩}.
    \label{tab:ignition_data}
\end{table*}

We performed the sensitivity study across a grid of astrophysical conditions representative of Type I X-ray burst sources (\cite{Galloway_2020}). We varied the accreted hydrogen mass fraction from $X_{H,\rm acc}=0.1$ (typical of helium-rich sources such as 4U 1820–303) up to $0.7$ (representative of hydrogen-rich “clocked bursters” like GS 1826–24), metallicity from $Z=0.01$ to $0.02$, and accretion rate from $\dot m=0.05$ to $0.9,\dot m_{\rm Edd}$. These ranges encompass the diversity of system parameters inferred from observational studies of X-ray bursters \cite{2017PASA...34...19G}. The values of $\dot m$, $X_{H,\rm acc}$, and $Z$ lie within the observationally inferred ranges for X-ray burst systems, as supported by model–observation studies (\cite{Cumming_2003}, \cite{Galloway_2008}, \cite{10.1093/mnras/stae2422}, \cite{10.1093/pasj/psae117}, \cite{takeda2025ninjasatmonitoringtypeixray} ). Table \ref{tab:ignition_data} summarizes the specific set of accretion conditions and ignition parameters resulting from our grid. We exclude pure He accretion as we did not find a thermal instability using our methodology and chosen parameter ranges. However, several of our models have very low hydrogen fractions at ignition, capturing the nuclear physics of He-dominated bursts.


SETTLE (\cite{Cumming_2000}) is a semi‐analytic model that calculates the thermal evolution of an accreted envelope on a neutron star up to the depth where a Type I X‐ray burst would ignite. SETTLE assumes steady-state hydrogen burning via the $\beta$‐limited hot CNO cycle, includes a constant base flux from the deeper stellar layers, and neglects the influence on previous bursts, the so-called thermal and compositional inertia (\cite{Woosley_2004}). The burst ignition condition is met when $\frac{d\epsilon_{3\alpha}}{dT}$ \textgreater $\frac{d\epsilon_{cool}}{dT}$, where $\epsilon_{3\alpha}$ is the energy production rate due to the $3\alpha$ reaction and $\epsilon_{cool}$ is the radiative cooling rate. The model is parameterized by the hydrogen mass fraction ($X_{\rm H,acc}$), the mass fraction of CNO nuclei ($Z_{\rm CNO}$) of the accreted material, the mass accretion rate ($\dot m$) scaled to the Eddington mass accretion rate ($\dot m_{\rm Edd}=8.8 \times 10^{4}\, g\, cm^{-2}\, s^{-1}$) and a constant base heating flux of $0.15$~MeV nucleon$^{-1}$ from the neutron star interior.

The original version of SETTLE assumes a composition of only hydrogen, helium, and CNO nuclei and thus $Z=Z_{\rm CNO}$. We modified the metallicity treatment in the SETTLE code by partitioning the metal fraction into an active component for the  CNO cycle and inert metals that do not contribute to nuclear heating. We use the total metallicity $Z$ of the accreted material as a parameter in our study, and set $Z_{\mathrm{CNO}} = 0.65 Z$ in SETTLE. The scaling factor is the sum of the mass fractions of catalysts of the hot CNO-1 cycle ($^{12}\mathrm{C}$, $^{13\text{--}15}\mathrm{N}$, and $^{14\text{--}15}\mathrm{O}$). We determined it at an early stage of post-processing of a SETTLE trajectory with NucNet Tools once CNO abundances are in equilibrium, but before additional CNO nuclei are produced by the 3-$\alpha$ reaction. The scaling factor agrees well with the mass fraction of CNO nuclei in the accreted solar composition.

The energy production rate with the scaled metal content in the SETTLE code follows:
\begin{equation}
    \epsilon_\mathrm{H} = 5.8 \times 10^{13} \left(\frac{Z_{\mathrm{CNO}}}{0.01}\right) \,\mathrm{erg} \, \mathrm{g}^{-1} \, \mathrm{s}^{-1}
\end{equation}
The hydrogen  depletion depth is calculated from \cite{Cumming_2000}:
\begin{equation}
    y_\mathrm{d} = 6.8 \times 10^{8} \,\mathrm{g} \, \mathrm{cm}^{-2} \left(\frac{\dot{m}}{0.1\dot{m}_{\mathrm{Edd}}}\right) \left(\frac{0.01}{Z_{\mathrm{CNO}}}\right) \left(\frac{X_{0}}{0.71}\right)
\end{equation}

\begin{figure*}[t] 
    \centering
    \begin{minipage}{0.46\textwidth}
        \centering
        \includegraphics[width=\linewidth]{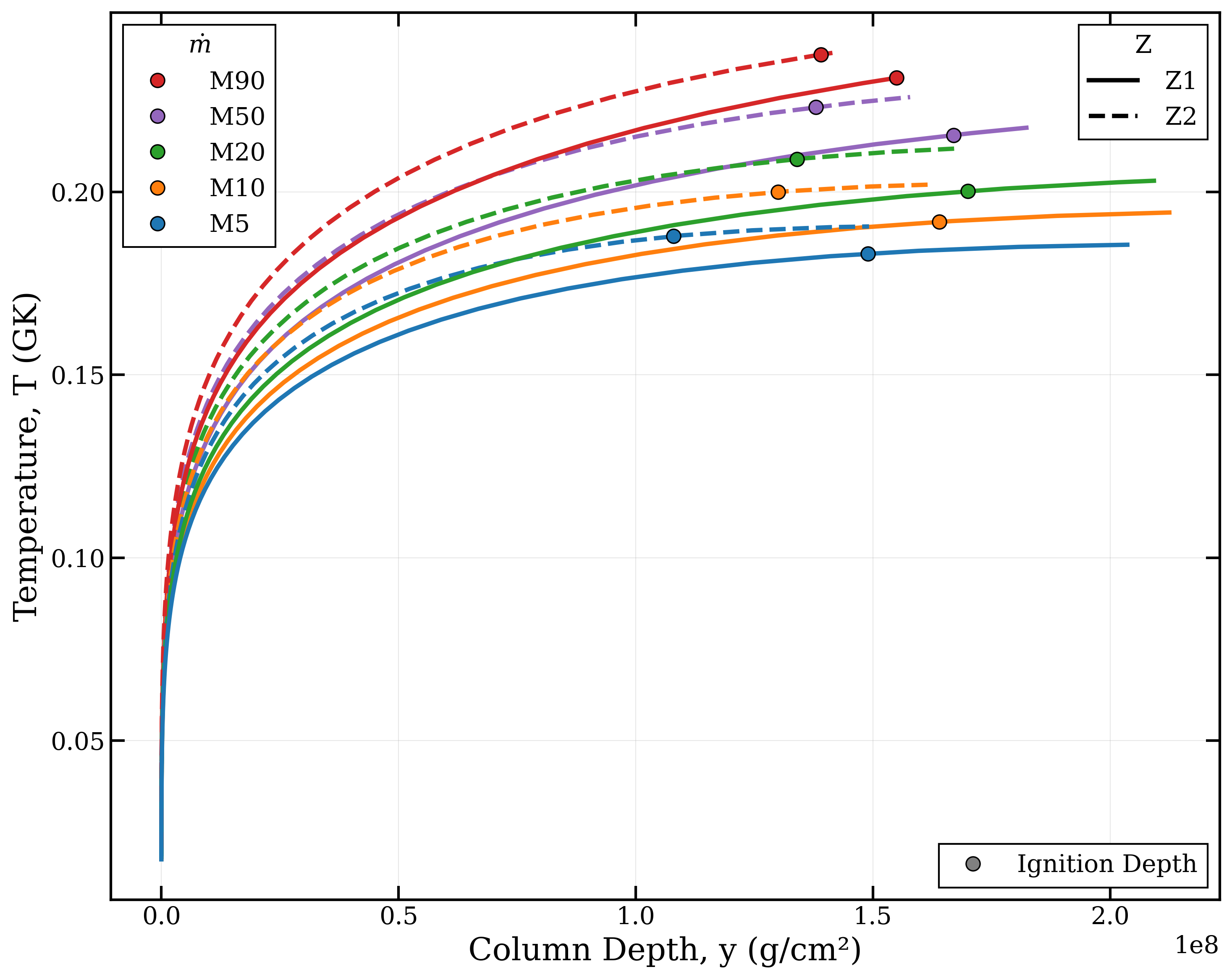}
        \vspace{2mm}
        \textbf{(a)}
    \end{minipage}
    \hfill
    \begin{minipage}{0.46\textwidth}
        \centering
        \includegraphics[width=\linewidth]{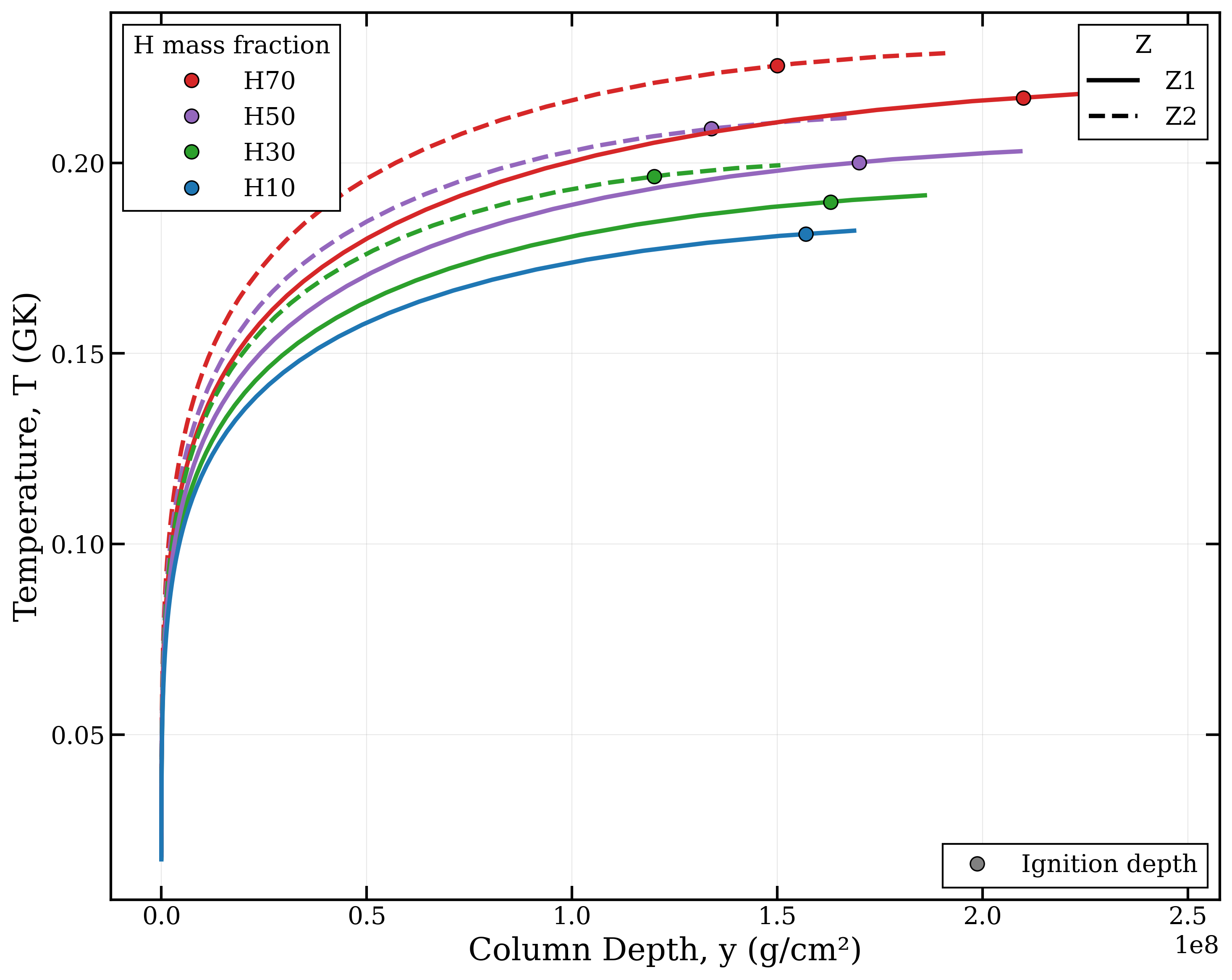}
        \vspace{2mm}
        \textbf{(b)}
    \end{minipage}
    
    \vspace{0.2cm} 
    \begin{minipage}{0.46\textwidth}
        \centering
        \includegraphics[width=\linewidth]{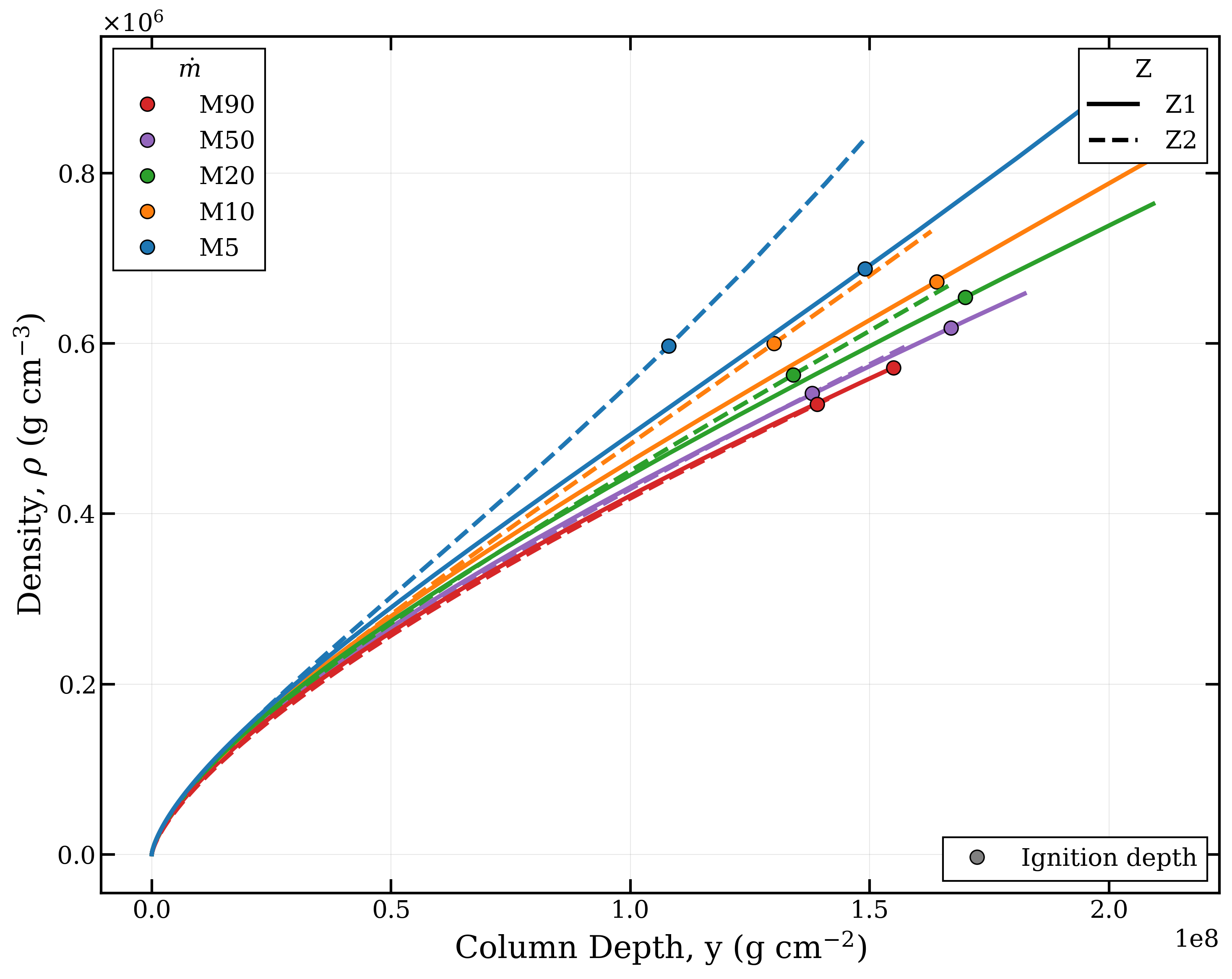}
        \vspace{2mm}
        \textbf{(c)}
    \end{minipage}
    \hfill
    \begin{minipage}{0.46\textwidth}
        \centering
        \includegraphics[width=\linewidth]{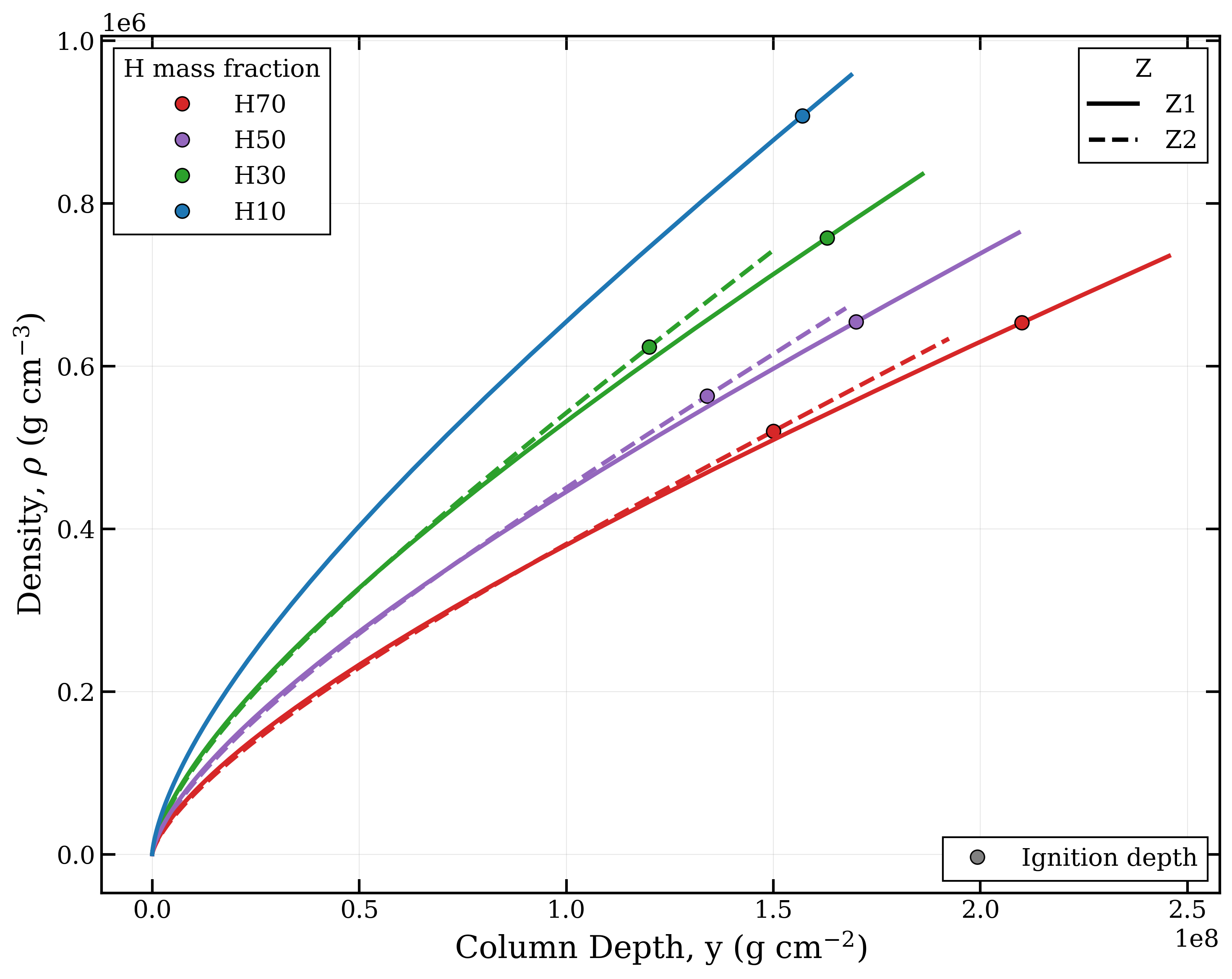}
        \vspace{2mm}
        \textbf{(d)}
    \end{minipage}
    \caption{Temperature and density profiles of the accreted envelope as functions of column depth. (a) Temperature profiles for fixed hydrogen mass fraction ($X_{\rm H,acc}=0.5$) while varying accretion rate ($\dot m$) and metallicity ($Z$). (b) Temperature profiles at fixed accretion rate ($\dot m = 0.2,\dot m_{\rm Edd}$) for different $X_{\rm H,acc}$ and $Z$. (c) Corresponding density profiles for the same cases shown in panel (a). (d) Corresponding density profiles for the cases in panel (b). Refer to Table \ref{tab:ignition_data} for the system acronym and parameter space values. The circles indicate the ONEZONE ignition depth for each system following the process described in the text, while the endpoint of each profile corresponds to the SETTLE ignition depth.}
    \label{fig:trajectory}
\end{figure*}

For each of the 32 systems in our grid of LMXBs' parameters, we used the modified SETTLE code to generate temperature and density profiles, integrating the envelope down to the depth of XRBs' ignition.  Figure \ref{fig:trajectory} shows these profiles for representative combinations of hydrogen mass fraction, metallicity, and accretion rate. At a given depth, systems with higher $\dot{m}$ and larger $X_\mathrm{H,acc}$ exhibit systematically higher temperatures and lower densities compared to lower $\dot{m}$ and hydrogen–poor counterparts.  Increasing $Z$ raises both the temperature and density uniformly throughout the accumulating layer.

\begin{figure*}[t] 
    \centering
    \begin{minipage}{0.46\textwidth}
        \centering
        \includegraphics[width=\linewidth]{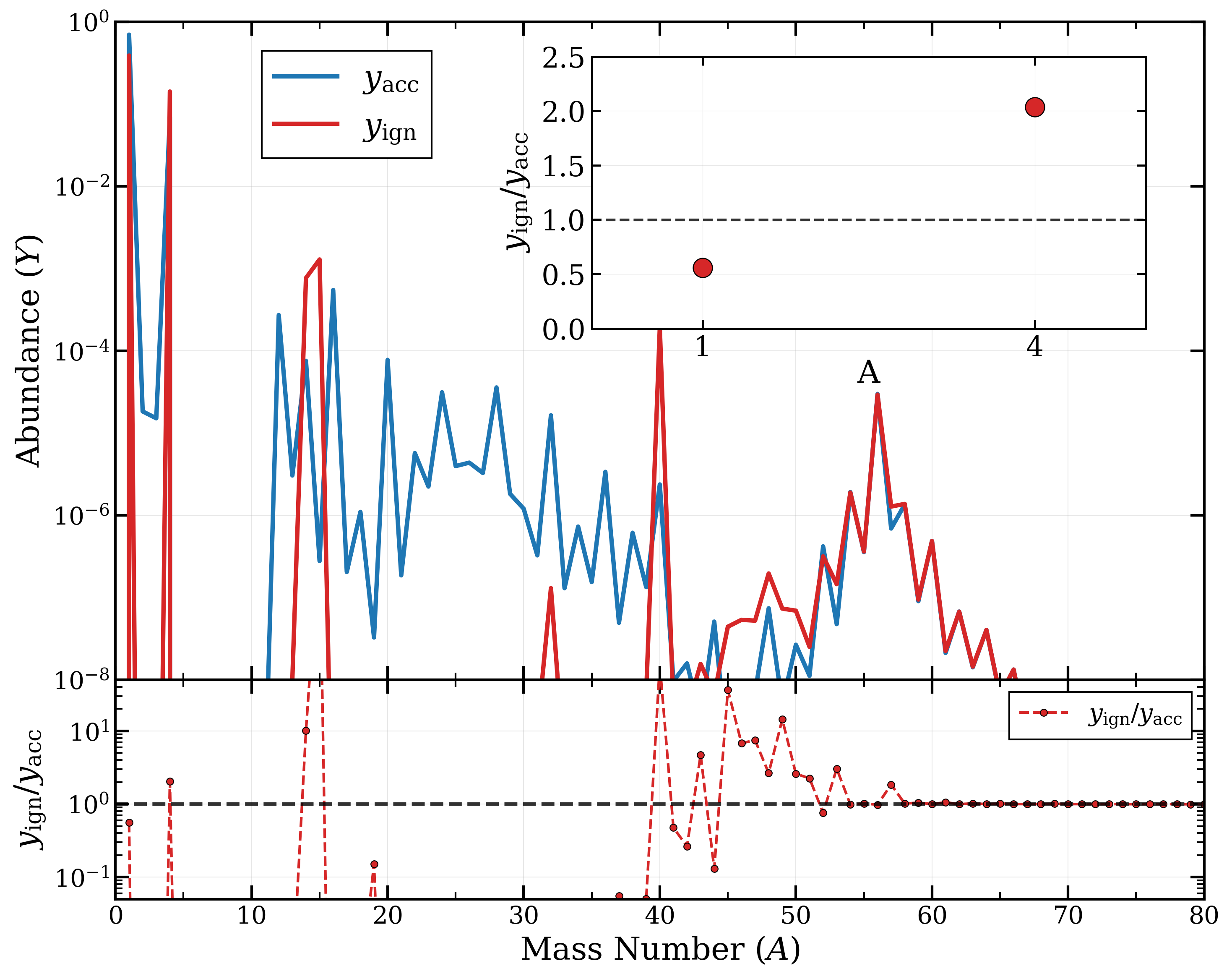}
        \vspace{2mm}
        \textbf{(a)}
    \end{minipage}
    \hspace{0.02\textwidth} 
    \begin{minipage}{0.46\textwidth}
        \centering
        \includegraphics[width=\linewidth]{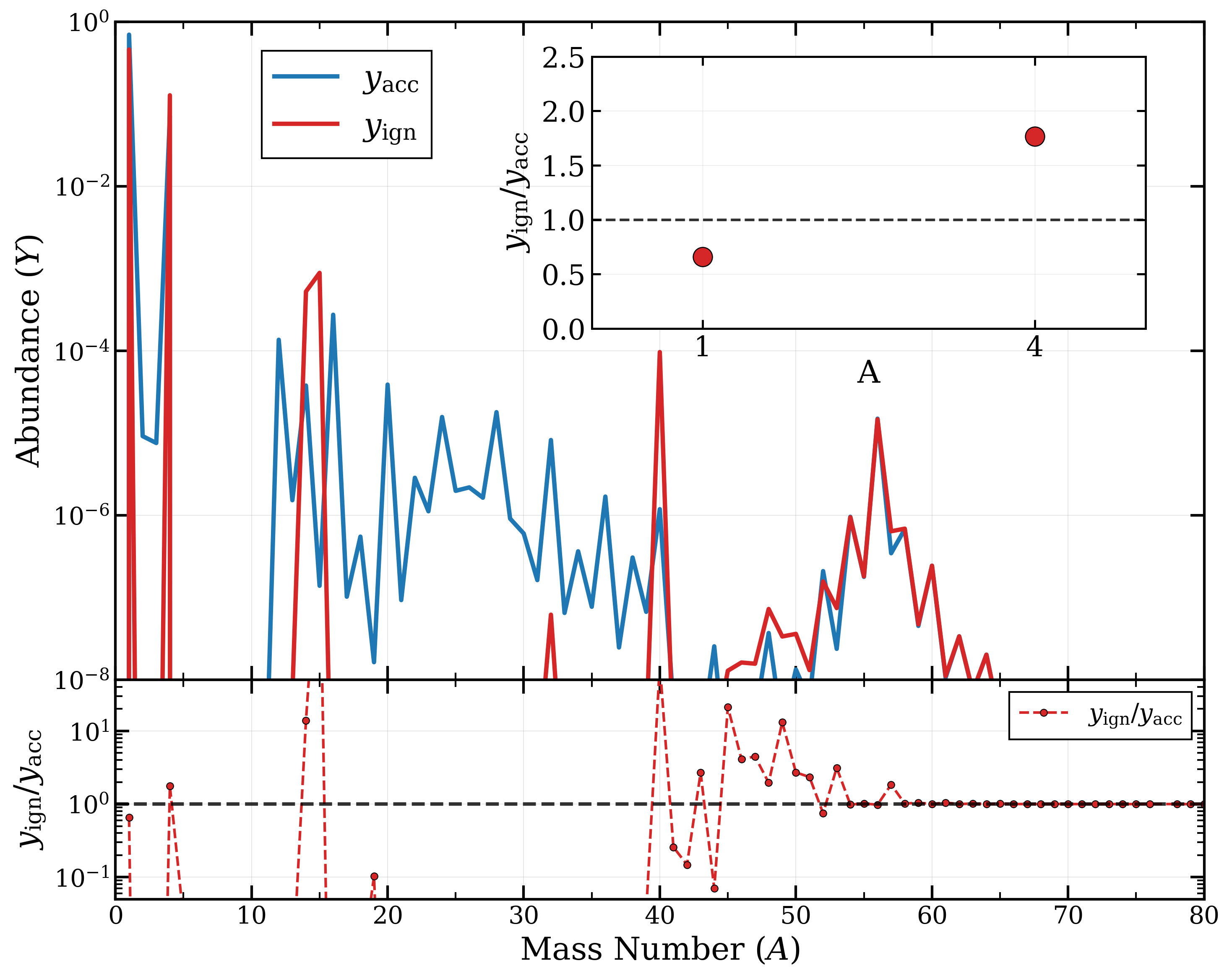}
        \vspace{2mm}
        \textbf{(b)}
    \end{minipage}
    
    \vspace{0.2cm} 
    \begin{minipage}{0.46\textwidth}
        \centering
        \includegraphics[width=\linewidth]{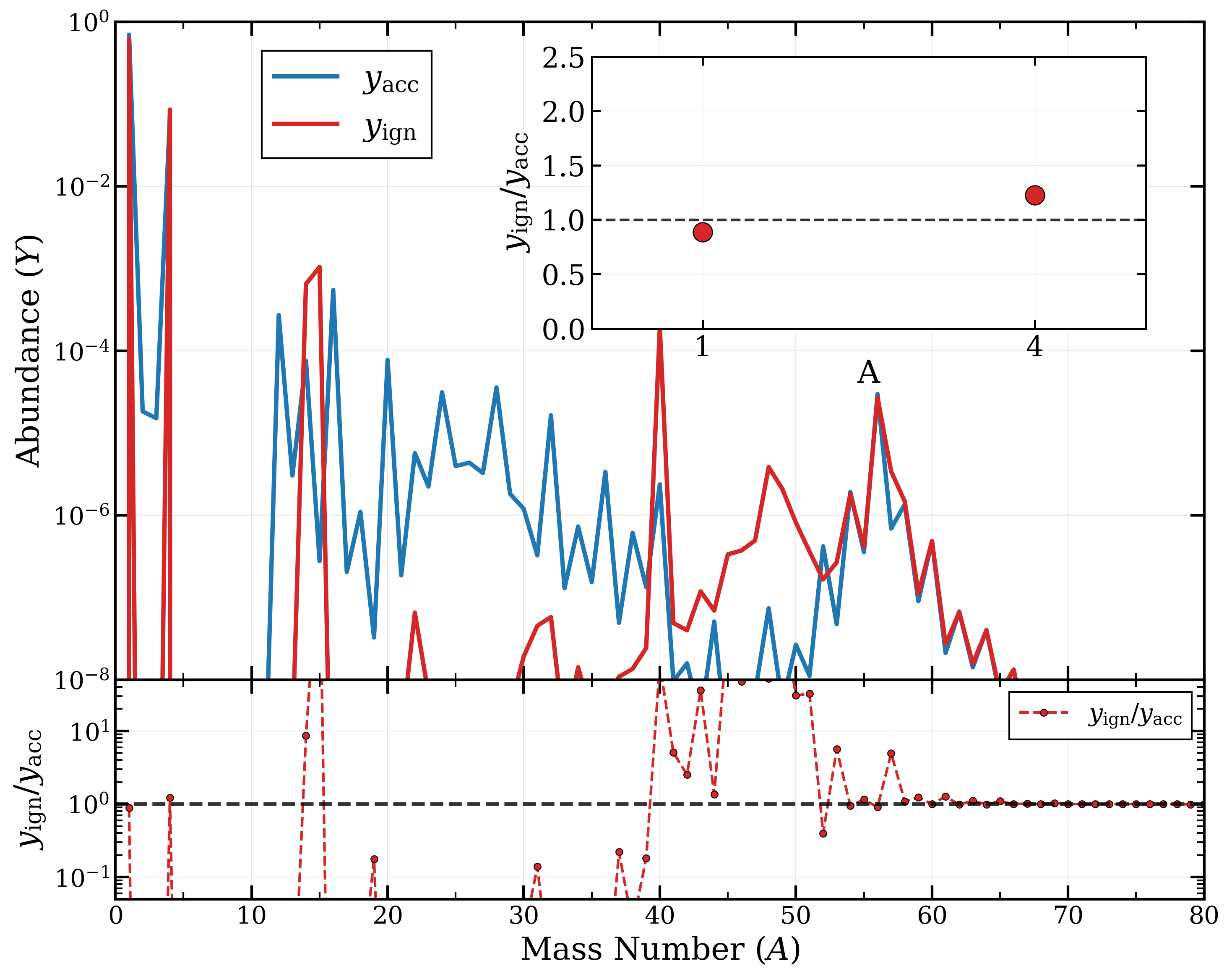}
        \vspace{2mm}
        \textbf{(c)}
    \end{minipage}
    \hspace{0.02\textwidth} 
    \begin{minipage}{0.46\textwidth}
        \centering
        \includegraphics[width=\linewidth]{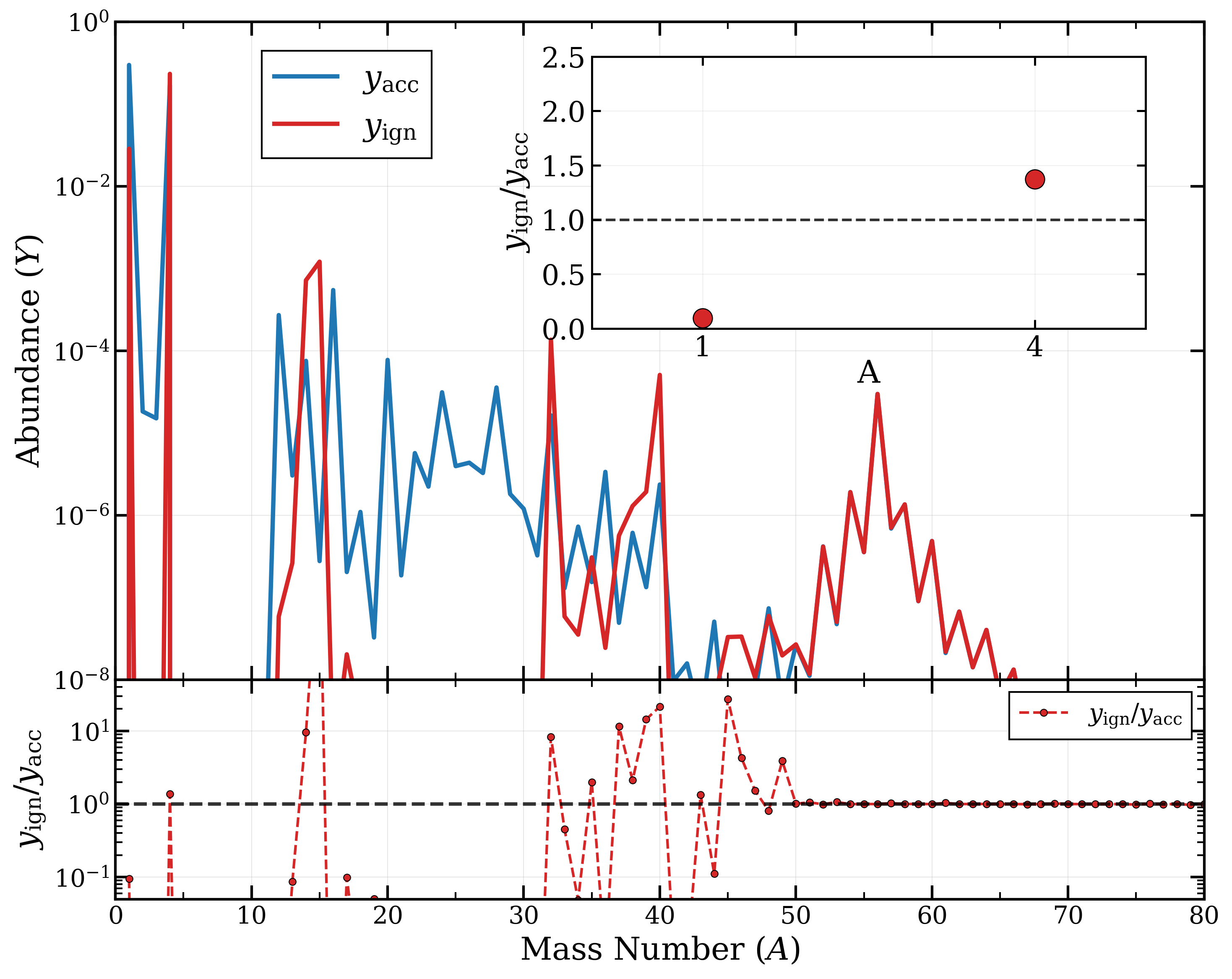}
        \vspace{2mm}
        \textbf{(d)}
    \end{minipage}
    \caption{Nuclear composition of the accreted material at the neutron star surface (blue line) and at ignition depth (red line), after the settling phase simulated using the NucNet
Tools single-zone reaction network code. The lower panel in each sub-figure shows the abundance ratio in each isotopic abundance between ignition and accretion depth. Binary system parameters are as follows: 
(a) $X_\mathrm{H,acc}= 0.7$, Z=0.02, and $\dot m=0.1 \dot m_\mathrm{Edd}$. (b) $X_\mathrm{H,acc}= 0.7$, Z=0.01, and $\dot m=0.1 \dot m_\mathrm{Edd}$ (c) $X_\mathrm{H,acc}= 0.7$, Z=0.02, and $\dot m=0.5 \dot m_\mathrm{Edd}$  (d) $X_\mathrm{H,acc}= 0.3$, Z=0.02, and $\dot m=0.1 \dot m_\mathrm{Edd}$. The insets in each panel show the abundance ratio ($y_\mathrm{ign}/y_\mathrm{acc}$) for highlighting the change in $^1$H and $^4$He prior to ignition.} 
    \label{fig:settling}
\end{figure*}

Nucleosynthetic processes modify the composition of the accreted material as it sinks from the neutron star surface up to the ignition depths. The details of the resulting isotopic distribution depend on the characteristics of the accretion process. 
 We used the NucNet Tools single‐zone reaction network \citep{Meyer:2013iM} to post-process each thermal trajectory produced by SETTLE and calculate the composition at ignition. NucNet Tools integrates a set of coupled differential equations for isotopic abundances under time‐dependent temperature and density conditions. The network spans nuclei from hydrogen to xenon, including isotopes along both the valley of stability and the proton drip line, thereby capturing the full range of $\alpha p$- and rp-process pathways relevant to XRB nucleosynthesis. All thermonuclear reaction rates are adopted from the JINA REACLIB v2.2 library \citep{2010ApJS..189..240C}. We initialized each network with the specified accreted hydrogen and helium mass fractions, then adjusted the residual metal fraction $Z$ by scaling the solar metallicity distribution of \cite{lodders2019solarelementalabundances}. Figure~\ref{fig:settling} shows the resulting isotopic distributions, both at the surface and the ignition depth, for representative accretion conditions. These profiles demonstrate distinct signatures of $Z$ and $\dot m$ on hydrogen burning; higher metallicity drives more efficient CNO burning and steeper hydrogen depletion, whereas larger accretion rates reduce the time available for steady hydrogen burning, resulting in a higher hydrogen fraction at ignition. Nuclear reactions prior to ignition also modify the initial heavy element distribution up to A$\sim$50 considerably. 

In our sensitivity study, we used a self-consistent single-zone reaction network code, the ONEZONE model (\cite{Cyburt_2016}), which dynamically evolves the thermodynamic conditions based on the energy generated during the burst. The general physics model adopted by the code is explained in \cite{1999A&A...342..464K}, \cite{Schatz_1999}, and \cite{2010ApJS..189..240C, Cyburt_2016}. The key details are summarized here. The ONEZONE code simulates the thermonuclear processes that occur during an XRB under specified ignition conditions. The code tracks the evolution of temperature, density, and composition within a single layer of the accreting system, assuming a constant pressure for the burning region. The model neglects detailed radiation transport, convection, and spatial gradient of thermal properties, focusing on the energy balance between nuclear heating, radiative cooling, and neutrino losses. We integrated the same network of 713 proton‑rich nuclei used with NucNet Tools, using reaction rates from the JINA reaction database REACLIBV2.2 (\cite{2010ApJS..189..240C}).



We first used the ONEZONE model to determine the depth at which the XRB ignites for each of the 32 systems of the LMXB parameter grid. The depth at which the nuclear burning becomes unstable, and the burst ignites, is defined as the point where the temperature derivative of the net heating exceeds the derivative for radiative cooling: $\frac{d}{dT}\bigl(\epsilon_{\rm nuc}-\epsilon_{\nu}\bigr)\;>\;\frac{d\epsilon_{\rm cool}}{dT}\,$. Here $\epsilon_{\rm nuc}$ represents nuclear heating, $\epsilon_{\nu}$ neutrino cooling and $\epsilon_{\rm cool}$ radiative cooling. 
By including both full nuclear energy generation and neutrino losses, this criterion extends the \cite{Cumming_2000} instability criteria, which solely depend on the heating rate from the triple-alpha reaction exceeding the cooling rate. Figure~\ref{fig:trajectory} compares the resulting ignition column depths with those predicted by SETTLE, showing that the semi‐analytic model systematically overestimates the instability depth. A similar trend was observed in previous comparative studies with the Kepler model (\cite{10.1093/mnras/stz2638}).

To perform the sensitivity study, we first perform a \emph{baseline} ONEZONE calculation for each of the 32 parameterized LMXB systems, taking nuclear reaction rates from the JINA REACLIB v2.2 library. Then, for each of these systems, we perform calculations individually varying each of the 2,744 ($p,\gamma$), ($\alpha,p$), and ($\alpha,\gamma$) reaction rates in our network. For each reaction, we perform two calculations, either increasing (\emph{UP} variation) or decreasing (\emph{DN} variation) the reaction rate for the forward and reverse reactions by a factor of 100.



To quantify changes in the burst light curve, we aligned at the onset of the burst the luminosity profiles of each calculation with a modified rate with that of the corresponding baseline model, using a low threshold for luminosity to define a consistent burst start. We have integrated the absolute difference between the two curves to quantify the impact of the reaction rate on the burst's light curve:

\begin{equation}
F_{\mathrm{lc}} = \int |L_{rate}(t) - L_{baseline}(t)|  dt
\label{eqFlc}\end{equation}


Here, $L_{\mathrm{rate}}(t)$ is the specific luminosity of the model with the modified reaction rate, and $L_{\mathrm{baseline}}(t)$ is the specific luminosity of the baseline model. A reaction is deemed impactful when its $F_{\mathrm{lc}}$ value exceeds $2\%$ of the baseline integral value for an individual system.

To measure compositional changes in the burst ashes, we calculate the maximum abundance ratio $F_{\mathrm{ash}}$ of an isotopic chain:
\begin{equation}
    F_{\mathrm{ash}} = \max \left(\frac{Y_{i,\mathrm{rate}}}{Y_{i,\mathrm{baseline}}}, \frac{Y_{i,\mathrm{baseline}}}{Y_{i,\mathrm{rate}}}\right)
\label{eqFy}\end{equation}

In our analysis, we only consider $F_{\rm ash}$ values for isotopes whose abundance in the baseline calculation exceeds $Y_{i,\mathrm{baseline}}\ge10^{-6}$. A reaction is deemed impactful if its $F_{\rm ash}$ value is five or above.




\section{Results} \label{sec:results}

\begin{figure*}[t] 
    \centering
    \begin{minipage}{0.48\textwidth}
        \centering
        \includegraphics[width=\linewidth, height=7cm]{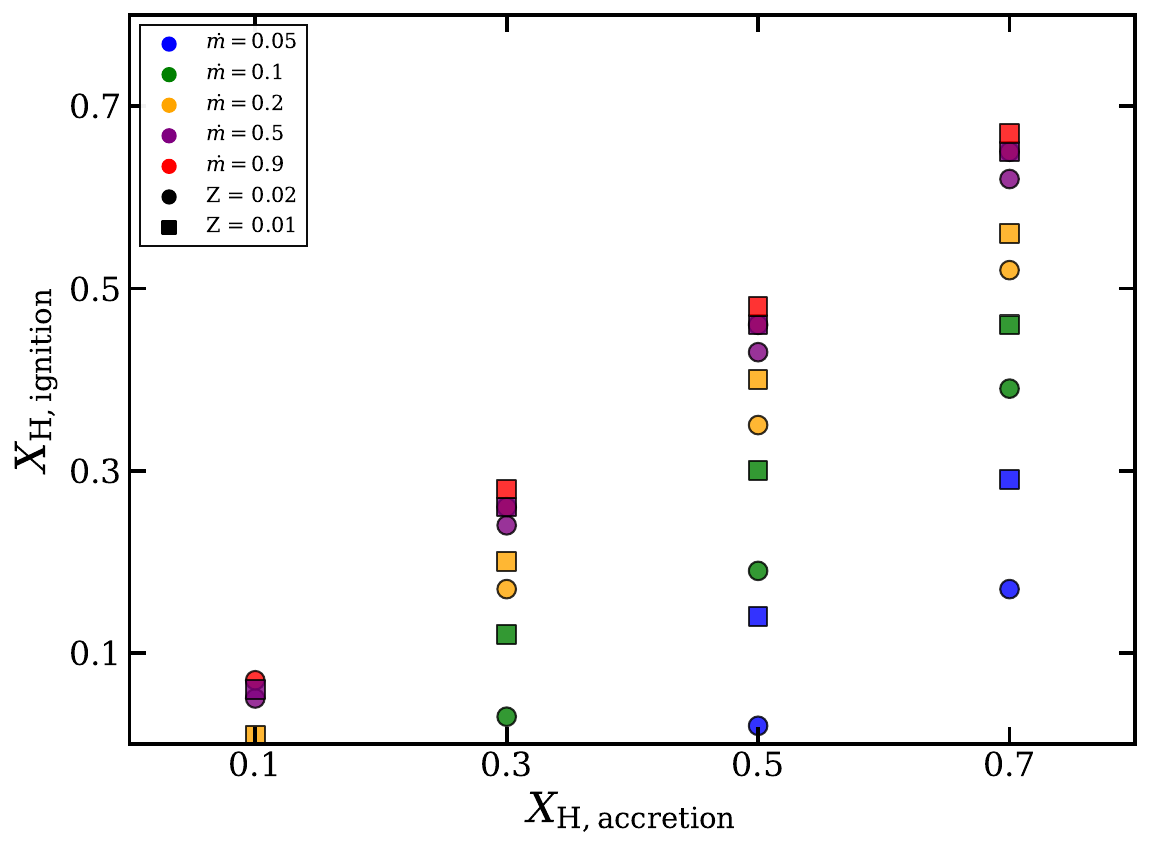}
        \vspace{2mm}
        \textbf{(a)}
    \end{minipage}
    \hfill
    \begin{minipage}{0.48\textwidth}
        \centering
        \includegraphics[width=\linewidth, height=7cm]{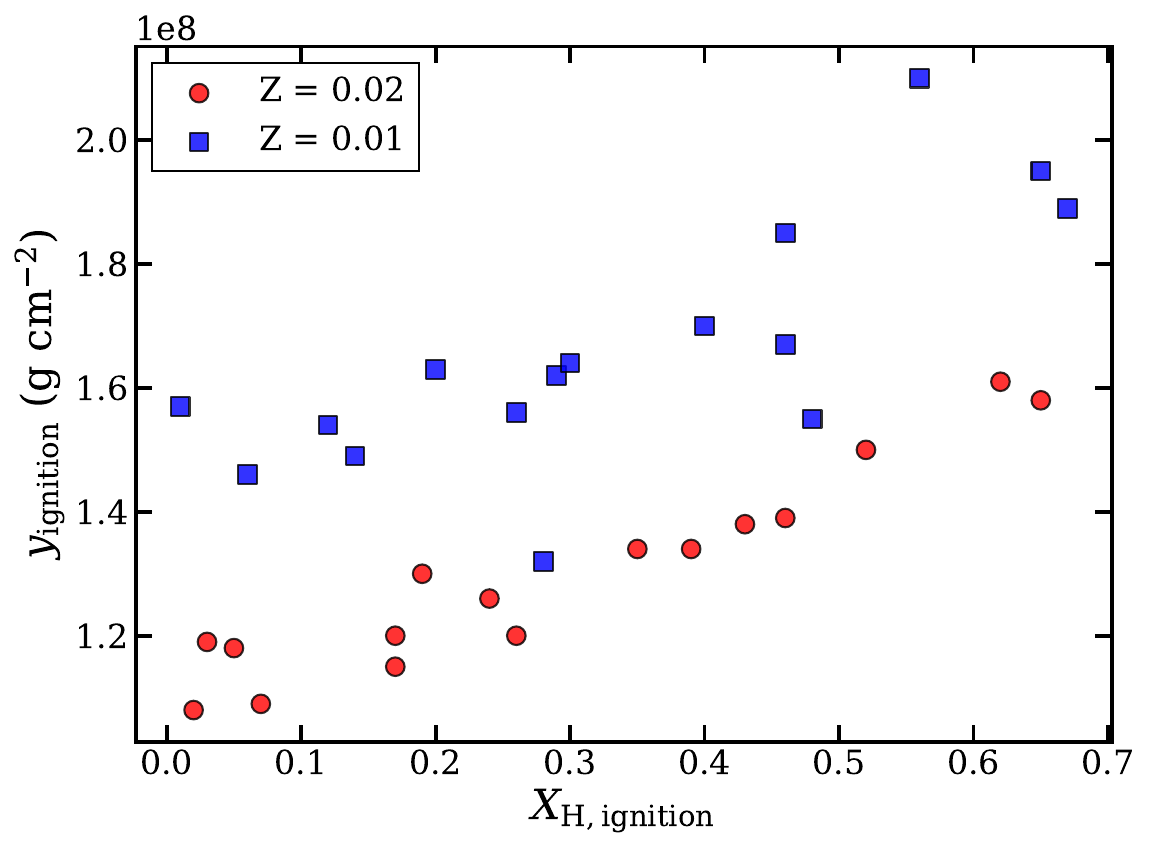}
        \vspace{2mm}
        \textbf{(b)}
    \end{minipage}
    \caption{(a) Hydrogen mass fraction at ignition ($X_{H,\rm ign}$) as a function of accreted hydrogen fraction ($X_{H,\rm acc}$) for various accretion rates ($\dot m$) and two metallicities. Circular markers indicate $Z = 0.02$, while square markers indicate $Z = 0.01$. Different colors correspond to different values of $\dot m / \dot m_{\rm Edd}$. (b) Ignition column depth vs hydrogen mass fraction at ignition for two metallicities.}
    \label{fig:binaryparms}
\end{figure*}

To explore how binary system parameters influence burst ignition conditions, we analyze the correlation between the accreted composition and ignition properties, as illustrated in Figure~\ref{fig:binaryparms}. The hydrogen mass fraction at ignition depends strongly on the initial accreted hydrogen content, with systematic variations across different accretion rates and metallicities. We find that systems with higher accretion rates and lower metallicities retain a larger fraction of unburned hydrogen at ignition, due to a shorter settling phase, which limits nuclear processing, and due to the slower CNO-cycle burning rate at lower $Z$, as shown in Figure~\ref{fig:binaryparms} (a). Figure~\ref{fig:binaryparms} (b) examines how the ignition column depth varies with the hydrogen mass fraction at ignition. We find that the models with a higher hydrogen content at ignition are correlated with igniting at deeper layers of the model trajectory. Also, a lower abundance of CNO catalysts not only reduces the rate of hydrogen burning via the CNO cycle but also pushes the ignition of the thermonuclear runaway deeper into the envelope.

\begin{figure*}[ht!]
\centering
\includegraphics[width=2.0\columnwidth]{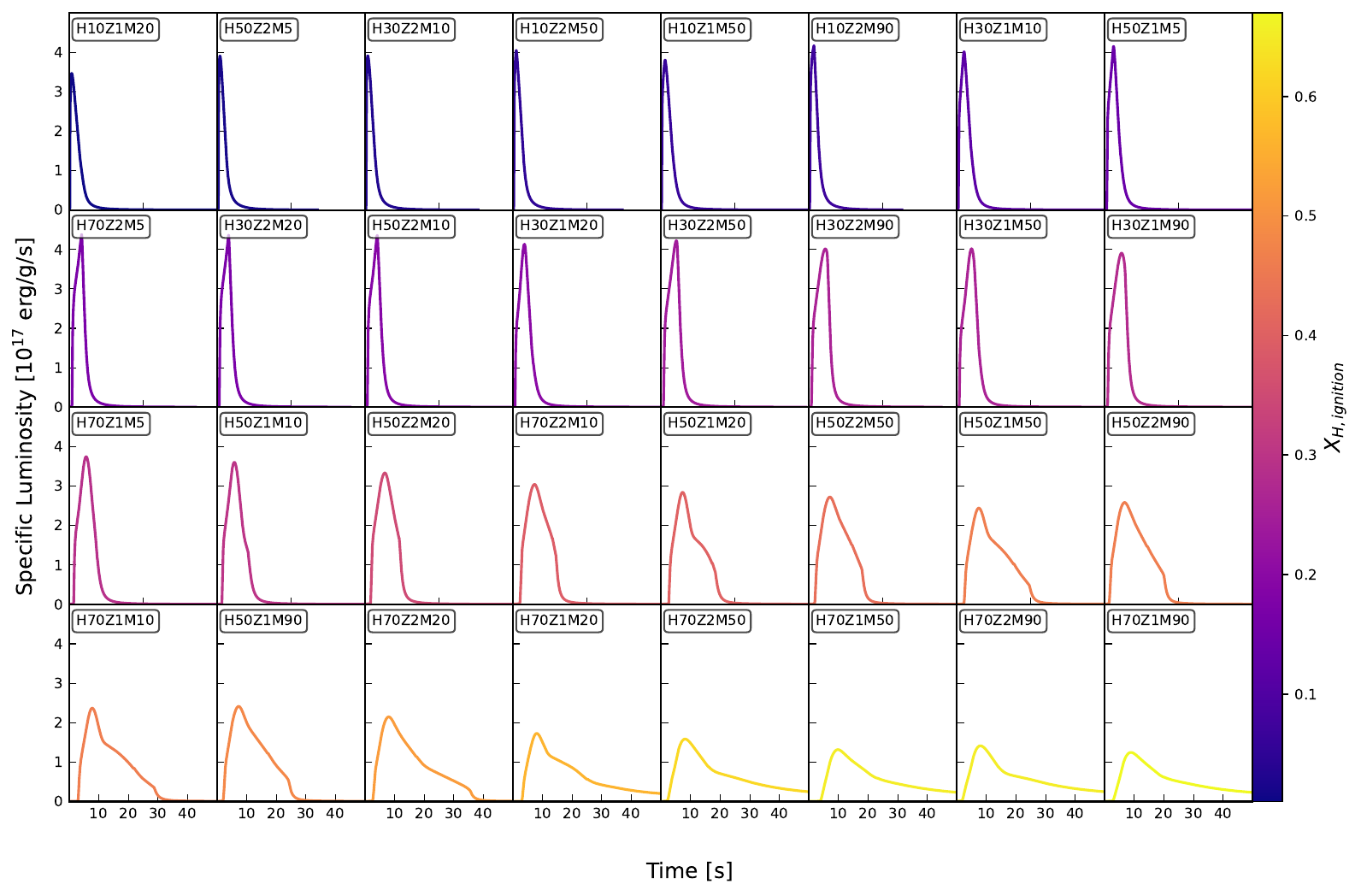} 
\caption{Baseline X-ray burst light curves for 32 systems. Panels are arranged so that $X_{H,\rm ign}$ decreases from the bottom-right toward the top-left, with the top row representing the most helium-rich cases. The color bar indicates the amount of hydrogen mass fraction at ignition ($X_{H,\rm ign}$).}
\label{fig:dasfigurinene_lc}
\end{figure*}

\begin{figure*}[ht!]
\centering
\includegraphics[width=2.0\columnwidth]{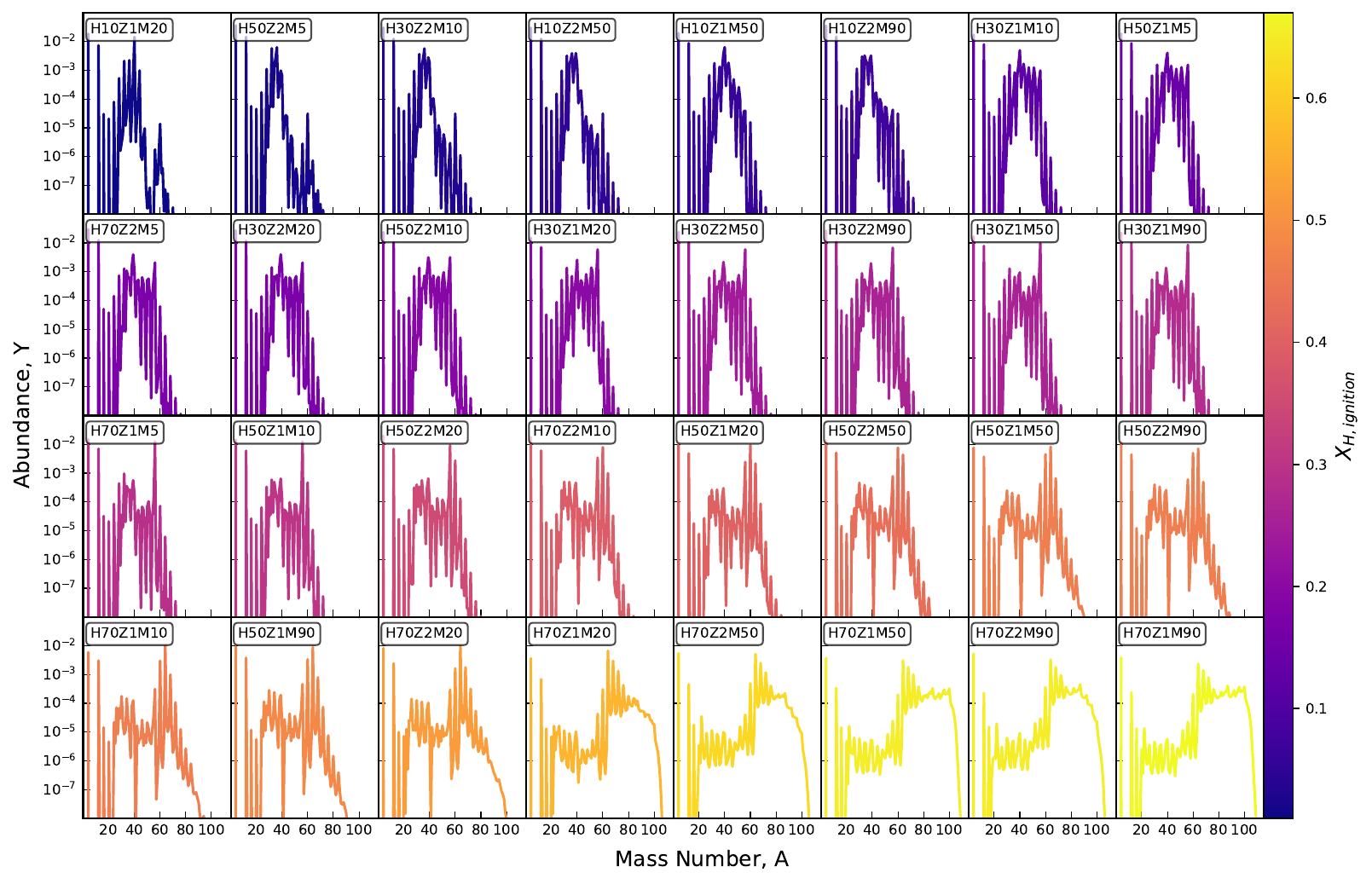} 
\caption{Baseline isotopic abundance distributions for the same 32 systems, using the identical panel ordering as Figure~\ref{fig:dasfigurinene_lc}.}
\label{fig:dasfigurinene_ash}
\end{figure*}

The baseline results for the 32 system parameter sets in our study are shown in Figures~\ref{fig:dasfigurinene_lc} and \ref{fig:dasfigurinene_ash}, illustrating the effect on the burst light curve and nucleosynthetic yields due to the variations in accreted composition and accretion rate. We see systems with higher hydrogen content at ignition tend to produce longer-duration bursts up to $\sim 200$ s due to prolonged hydrogen burning with an extended rp-process path. In contrast, He-rich systems ($X_\mathrm{H,\rm ign} \lesssim 0.3$) produce shorter duration bursts no longer than $\sim 15$ s due to rapid energy release from helium ignition without an extended rp-process tail. As with the light curves, the ignition composition shapes the abundance pattern. Mixed H/He bursts with moderate-to-high $X_\mathrm{H,\rm ign}$ tend to synthesize nuclei in the heavier mass range up to A $\sim 110$, characteristic of an extended rp-process path. In contrast, He-rich bursts terminate earlier in the reaction flow, with final abundances concentrated near A $\sim 60$ medium mass range. 

The list of reactions that were found to have a large impact on the results of the XRB model is summarized in tables Table~\ref{tab:ranked_sensitivity_lc} and ~\ref{tab:ranked_sensitivity_ash}. Table~\ref{tab:ranked_sensitivity_lc} lists the 49 reactions whose rate variations significantly change the burst light curve shape. Likewise, Table~\ref{tab:ranked_sensitivity_ash} summarizes the 182 reactions that alter at least one isotopic abundance by a factor of five or more.

\begin{figure*}[t] 
    \centering
    \begin{minipage}{0.49\textwidth}
        \centering
        \includegraphics[width=\linewidth]{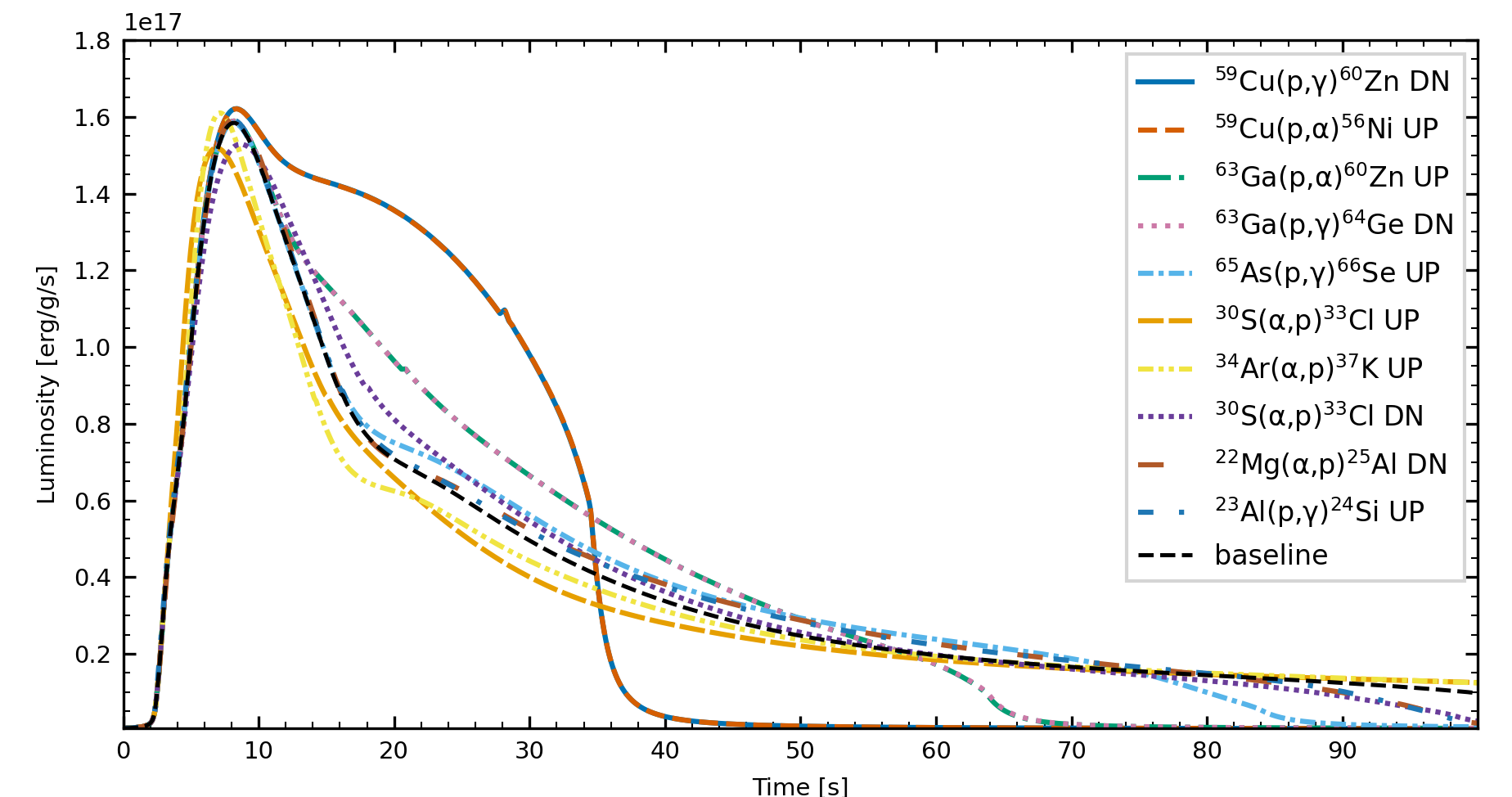}
    \end{minipage}
    \hfill
    \begin{minipage}{0.49\textwidth}
        \centering
        \includegraphics[width=\linewidth]{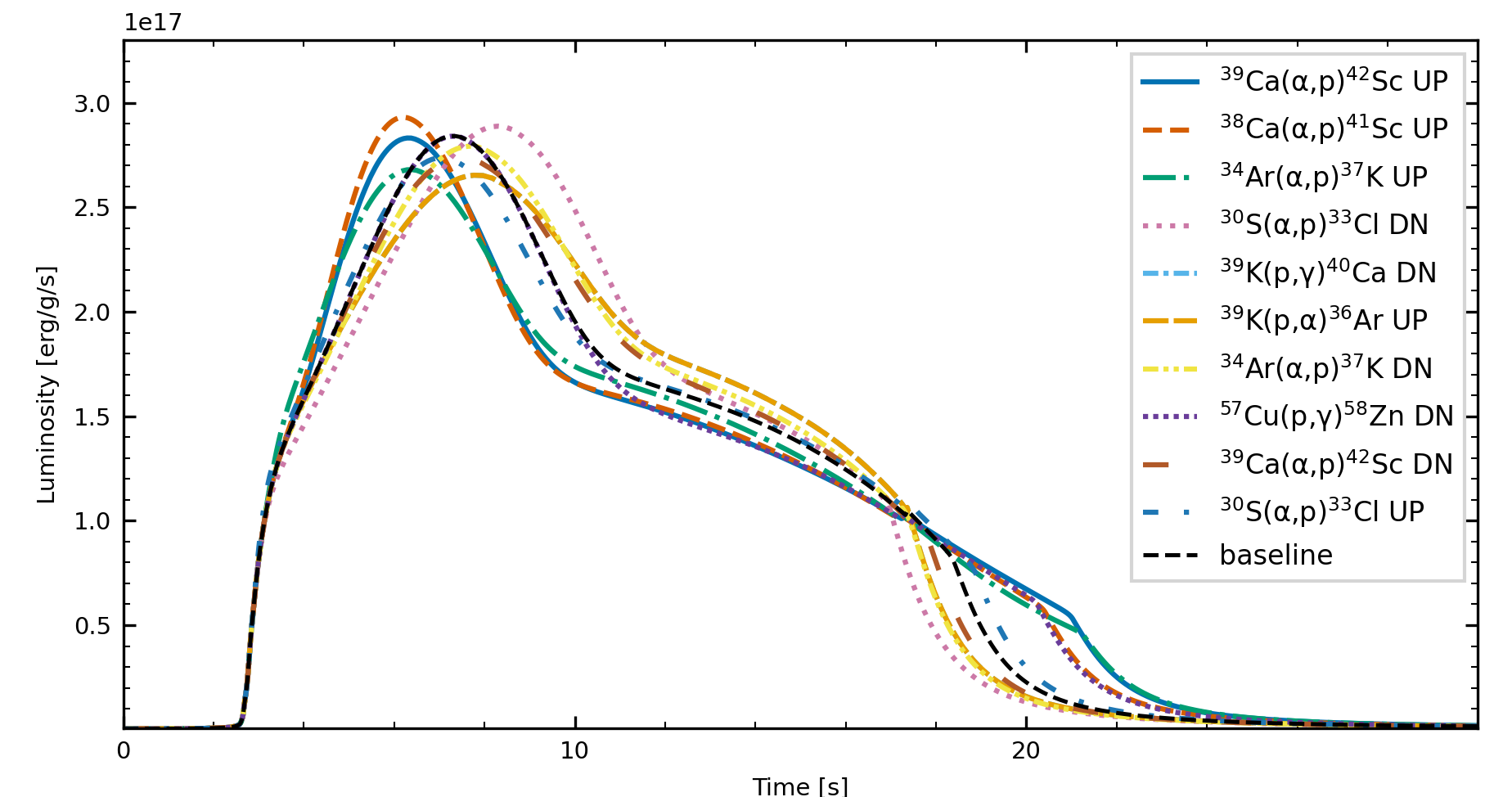}
    \end{minipage}
    
    \vspace{0.2cm} 

    \begin{minipage}{0.49\textwidth}
        \centering
        \includegraphics[width=\linewidth]{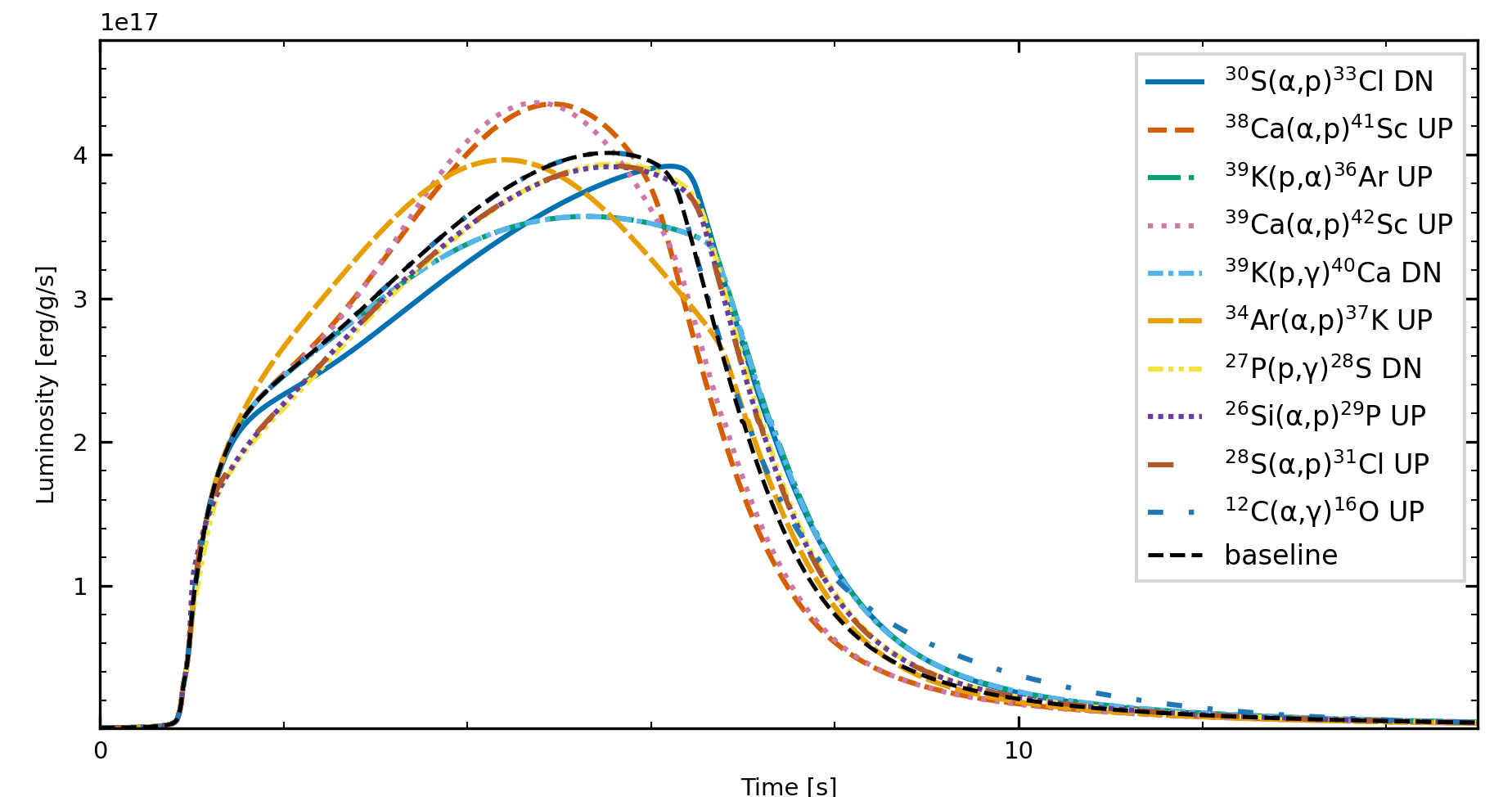}
    \end{minipage}
    \hfill
    \begin{minipage}{0.49\textwidth}
        \centering
        \includegraphics[width=\linewidth]{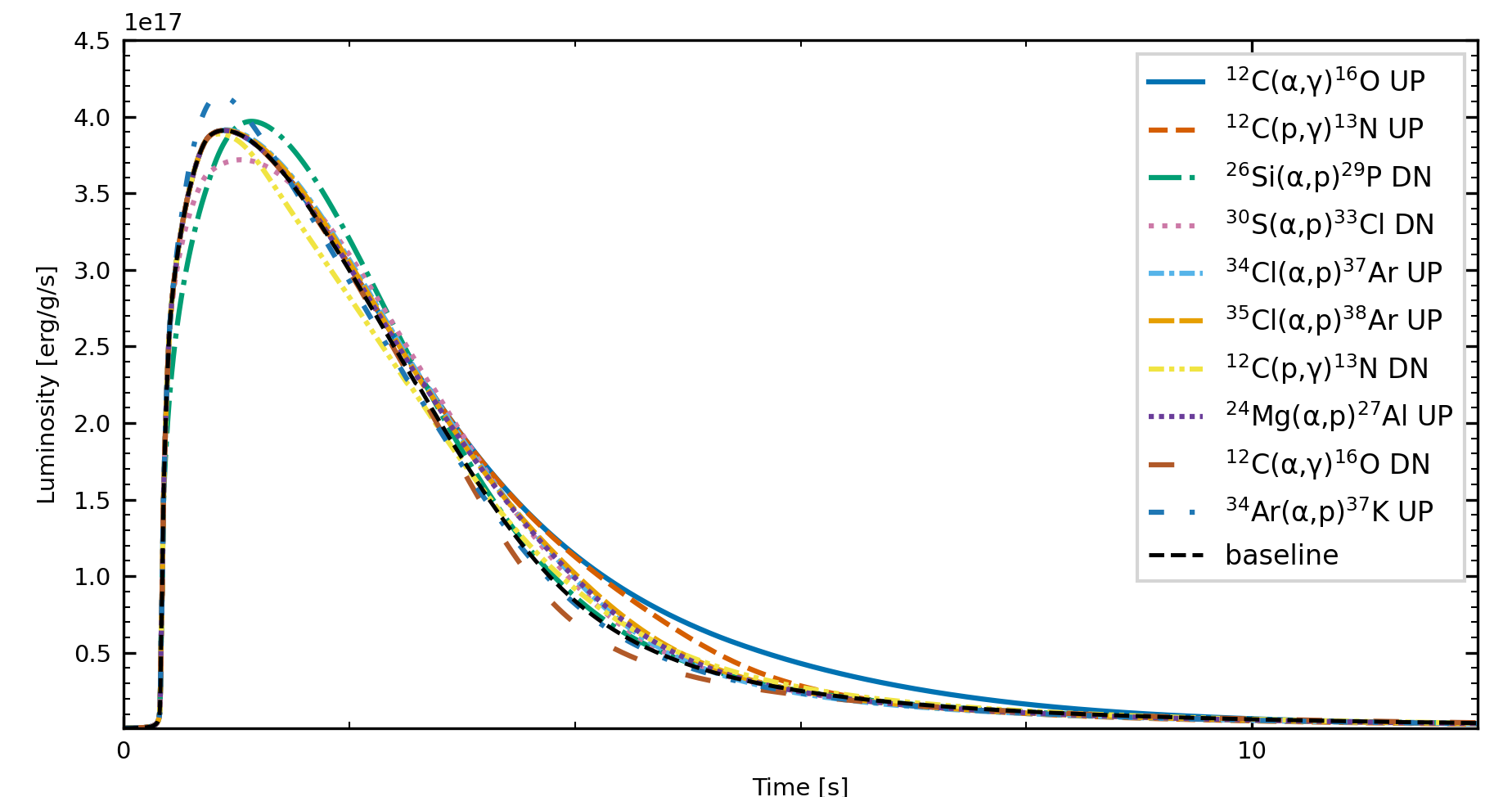}
    \end{minipage}

    \caption{Model light curves for the baseline calculation using the REACLIB v2.2 reaction rates (\emph{baseline}) and for calculations where individual reaction rates are varied by a factor of 100. The figure only shows results for the reaction rate variations that produced the 10 highest values for the light curve sensitivity factor (equation~\ref{eqFlc}). The four panels show results for binary systems with a variety of accretion parameters; from more hydrogen-rich to more helium-rich bursts, the models shown are H70Z2M50( top left), H50Z1M20 (top right), H30Z2M90 (bottom left), and H30Z2M10 (bottom right).  Note that more reactions exceed the sensitivity factor threshold as listed in Table~\ref{tab:ranked_sensitivity_lc}. }
    \label{fig:rep_group_lc}
\end{figure*}


\begin{figure*}[t] 
    \centering
    \begin{minipage}{0.49\textwidth}
        \centering
        \includegraphics[width=\linewidth]{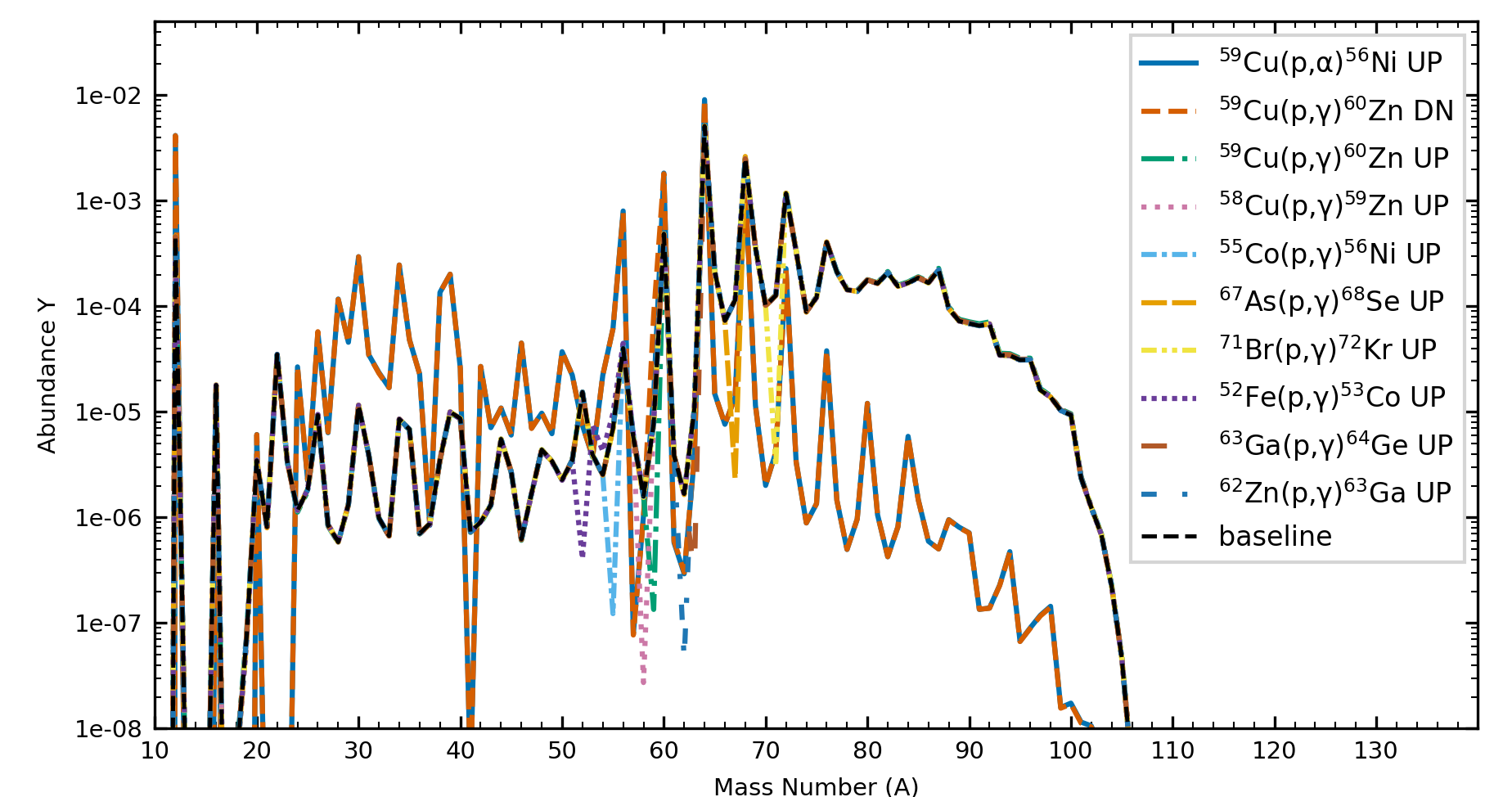}
    \end{minipage}
    \hfill
    \begin{minipage}{0.49\textwidth}
        \centering
        \includegraphics[width=\linewidth]{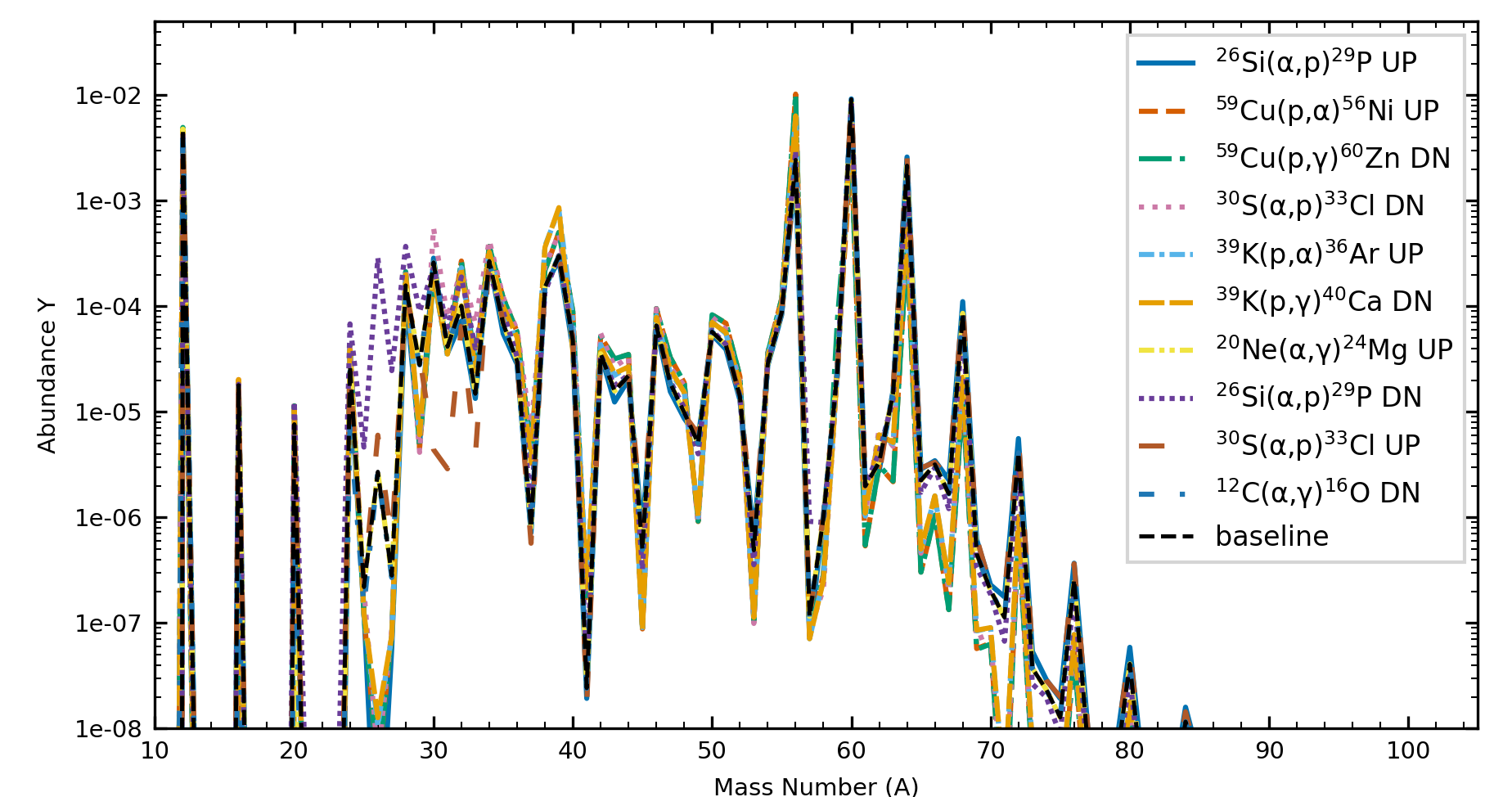}
    \end{minipage}
    
        \vspace{0.2cm} 

    \begin{minipage}{0.49\textwidth}
        \centering
        \includegraphics[width=\linewidth]{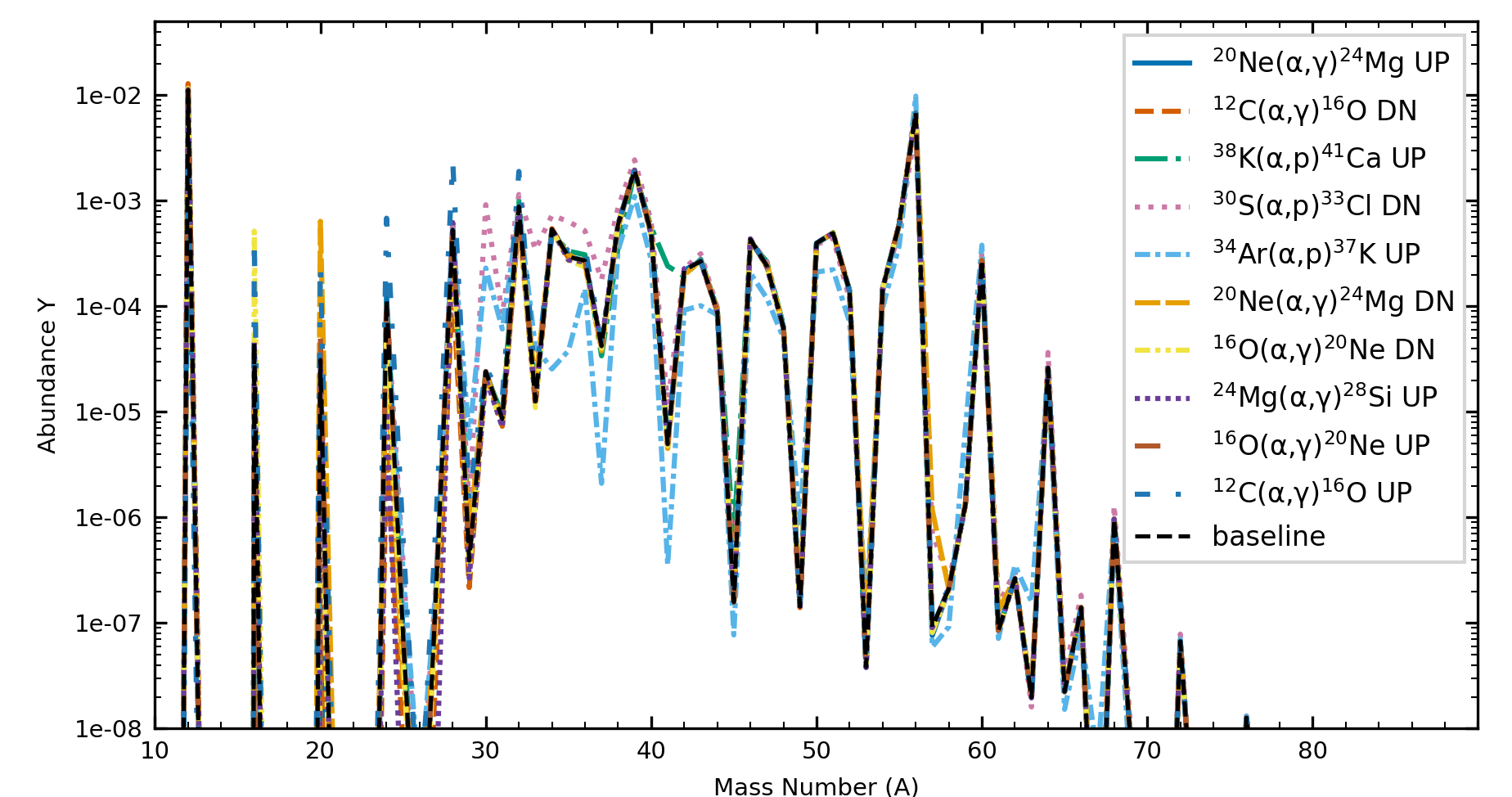}
    \end{minipage}
    \hfill
    \begin{minipage}{0.49\textwidth}
        \centering
        \includegraphics[width=\linewidth]{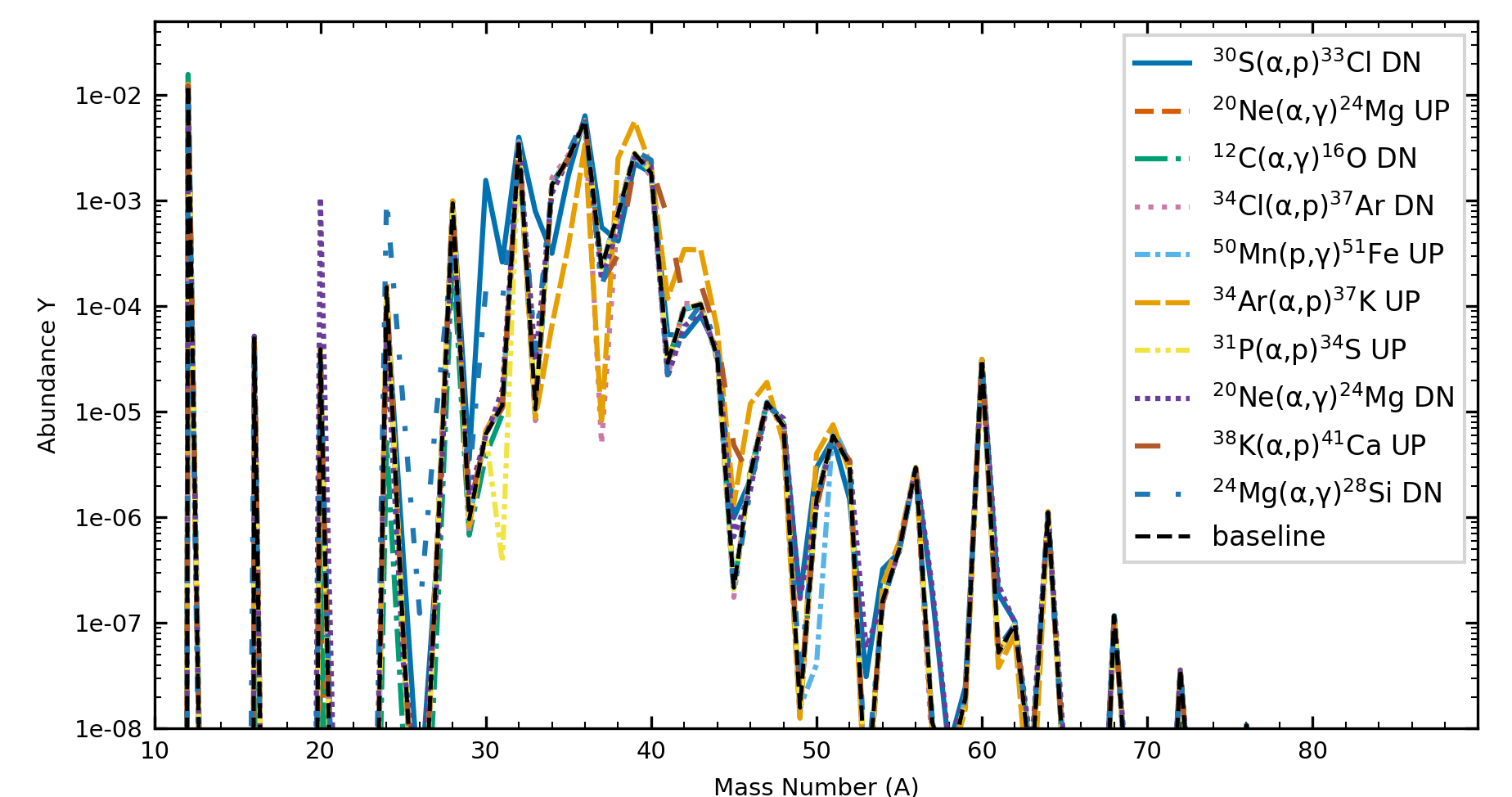}
    \end{minipage}

    \caption{Same as Figure~\ref{fig:rep_group_lc}. The impactful reaction rate variation changes the final abundances produced.}
    \label{fig:rep_group_ash}   
\end{figure*}

\begin{figure*}[!t]
    \centering
    \includegraphics[width=0.9\textwidth]{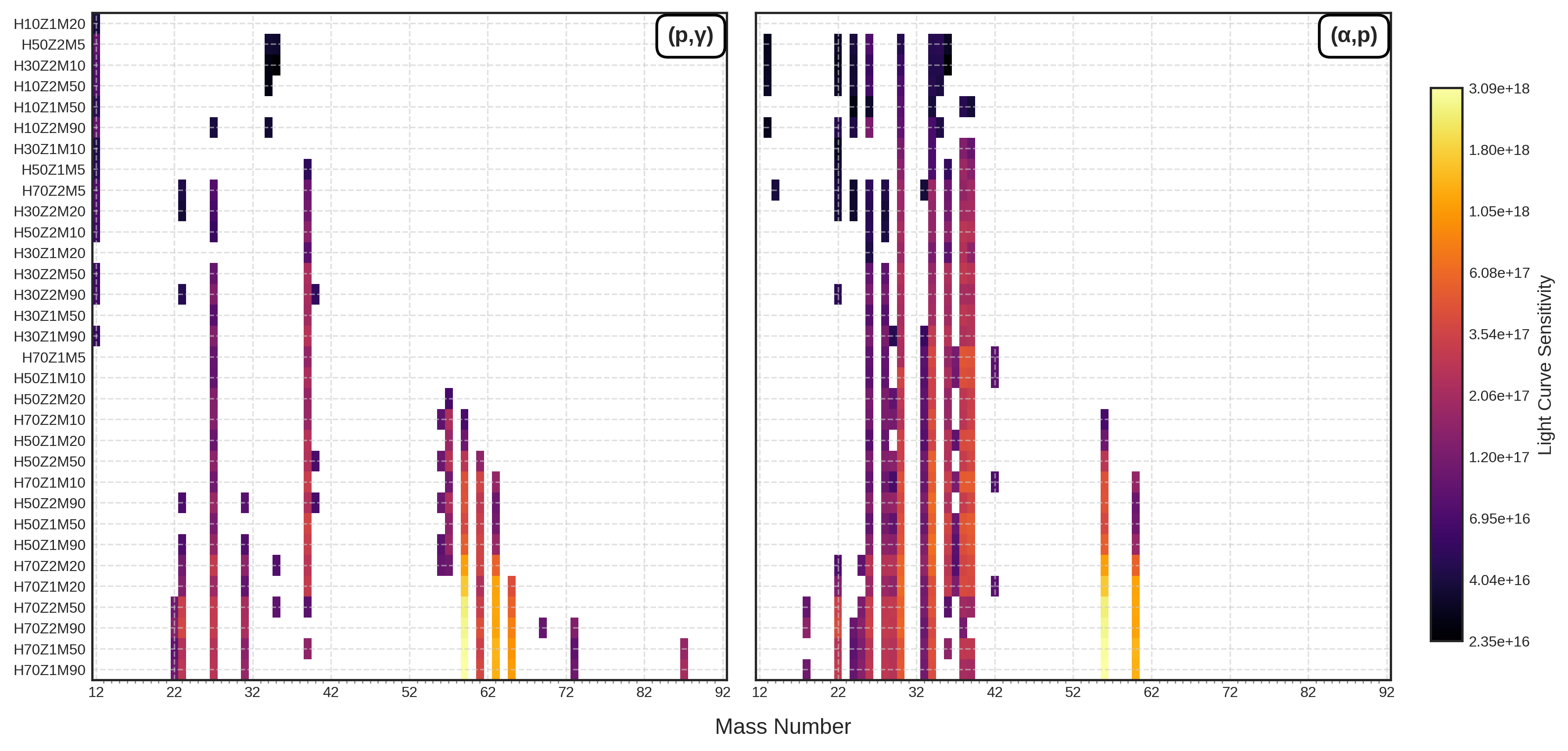}
    \vspace{0.6cm}
    \includegraphics[width=0.9\textwidth]{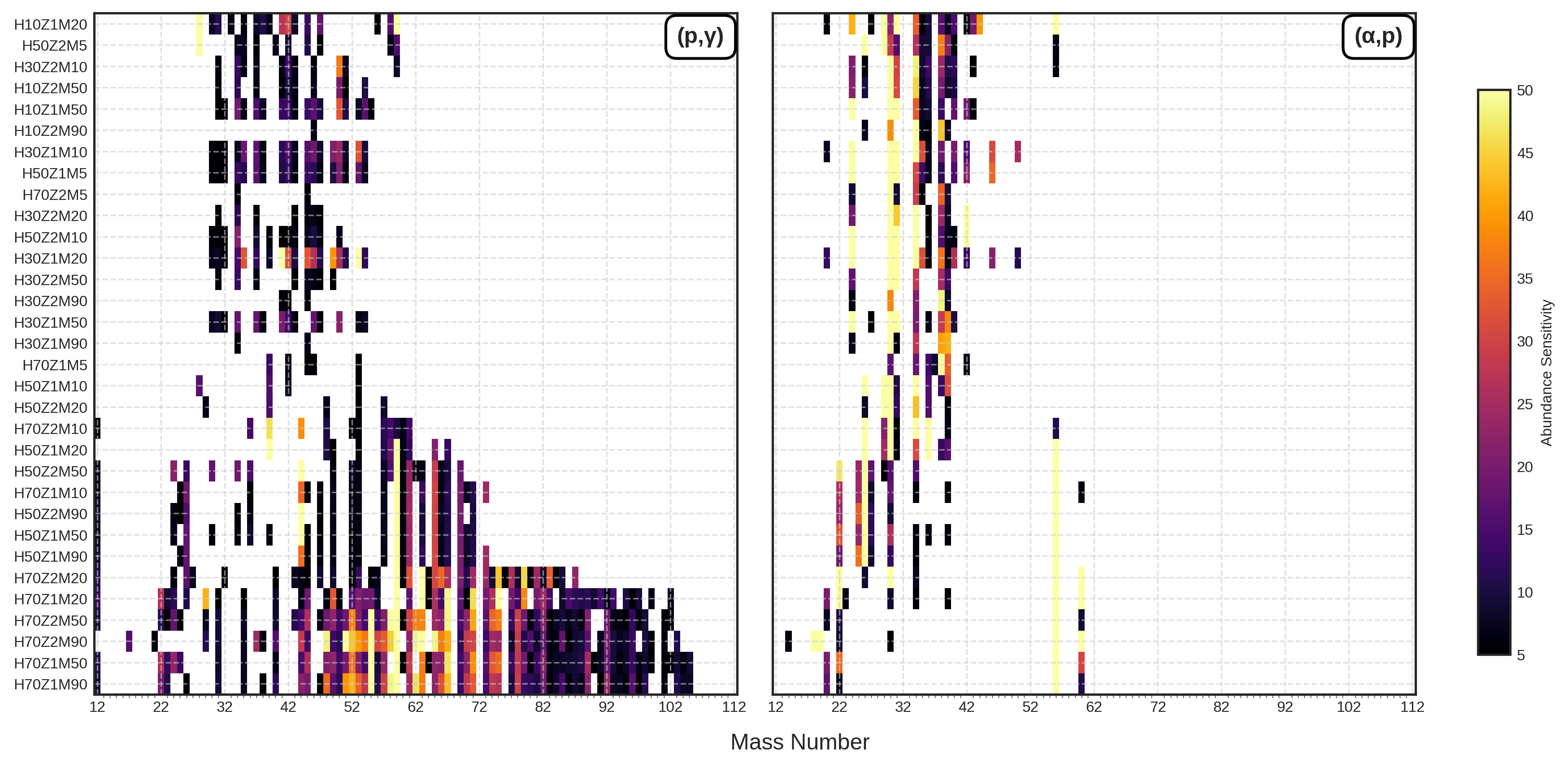}
    \caption{Visualization of the impact of individual nuclear reaction rates across all the 32 systems in our study. Each panel shows the impact factor as a function of the target nucleus's mass number and the model parameters. The models are ordered by $X_\mathrm{H, ign}$ (Table~\ref{tab:ignition_data}) with the most hydrogen-rich bursts placed at the bottom of the y-axis. For each mass number, we only include the reaction variation with the highest impact factor (see Table~\ref{tab:ranked_sensitivity_lc} and \ref{tab:ranked_sensitivity_ash}).  \textit{Top row:} Sensitivity factor for impact on the burst light curve (equation~\ref{eqFlc}), separated by ($p,\gamma$) and ($\alpha, p$) reactions (left and right panels, respectively). \textit{Bottom row:} Sensitivity factor for impact on the final abundances (equation~\ref{eqFy}), separated by ($p,\gamma$) and ($\alpha, p$) reactions (left and right panels, respectively). In Figures~\ref{fig:heatmap_combined}--\ref{fig:heatmap_lc_2}, the three $(p,\alpha)$ reactions ($^{39}$K$(p,\alpha)^{36}$Ar, $^{59}$Cu$(p,\alpha)^{56}$Ni, and $^{63}$Ga$(p,\alpha)^{60}$Zn) are represented using the inverse $(\alpha,p)$ notation. 
    } 
    \label{fig:heatmap_combined}
\end{figure*}

Figures~\ref{fig:rep_group_lc} and \ref{fig:rep_group_ash} illustrate the impact of the top-ranked reactions on the light curve and abundances obtained from the ONEZONE model, for four representative systems that span the full range of hydrogen mass fractions at ignition.  Figure~\ref{fig:rep_group_lc} illustrates how each rate variation changes the burst light‐curve shape, and Figure~\ref{fig:rep_group_ash} shows the corresponding shifts in the final abundance patterns due to variations in each reaction rate. The heatmaps in Figure~\ref{fig:heatmap_combined} show the systematic trends in reaction rate sensitivities across the full set of modeled systems. They illustrate the shift in dominant nuclear reaction pathways with varying ignition composition across 32 systems. In mixed H/He bursts ($X_\mathrm{H, ign} \gtrsim 0.5$), both $(p,\gamma)$ and $(\alpha,p)$ reactions contribute comparably to variations in the burst light curve, indicating that a mix of proton and alpha-capture pathways shapes energy generation in this regime. However, as $X_\mathrm{H, ign}$ decreases, $(\alpha,p)$ reactions increasingly dominate, especially in He-rich bursts, where they become the primary drivers of light curve variation while the $(p,\gamma)$ contributions diminish. A similar trend is observed in the final abundance sensitivity for the He-rich systems, where the $(\alpha,p)$ reactions surpass $(p,\gamma)$ in influence below $X_\mathrm{H, ign} \approx 0.5$. In contrast, the $(p,\gamma)$ reactions play a more prominent role in shaping the final composition in mixed H/He bursts ($X_\mathrm{H, ign} \gtrsim 0.5$). Together,  Figures~\ref{fig:dasfigurinene_lc}--\ref{fig:heatmap_combined} reveal an ignition composition dependent transition in dominant reaction pathways from $(p,\gamma)$ to $(\alpha,p)$ channels as $X_{H,\mathrm{ign}}$ varies, identifying a focused set of nuclear rates whose uncertainties have the largest impact across our parameter grid. Only a subset of reactions influences both the burst light curve and the final composition. There are 38 such reactions, which are listed in both Tables~\ref{tab:ranked_sensitivity_lc} and~\ref{tab:ranked_sensitivity_ash}. These overlap cases are dominated by bottleneck reactions that also govern the overall energy generation.

Among the most significant composition trends is the behavior of the $A=12$ mass chain. As shown in Figure~\ref{fig:rep_group_ash}, the final $^{12}$C abundance in the baseline model increases systematically from hydrogen‐rich to helium‐rich bursts. The largest mass fraction of $^{12}$C in the rp-process ashes is $\approx 18\%$ for model H10Z2M90. For helium‐rich systems with $X_{H,\mathrm{ign}} \approx 0.01$–0.28, we identify 12 reactions whose rate variations drive the largest increase in carbon production. In the models with these modified reaction rates, the final $^{12}$C mass fraction increased to values up to (10-23)$\%$. 

\begin{figure*}[!t]
    \centering
    \includegraphics[width=0.8\textwidth]{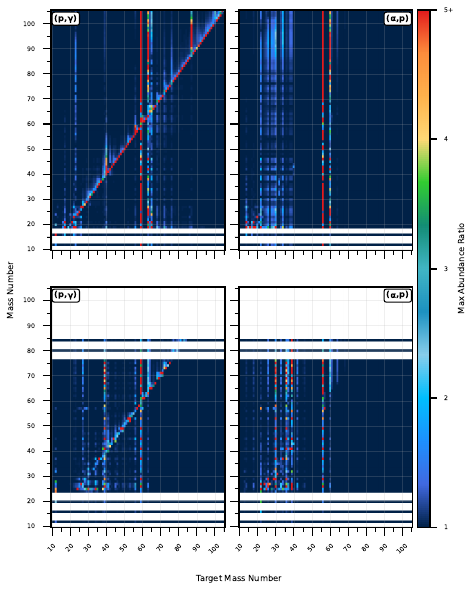}
    \caption{Mass number change as a function of target mass of the reaction nuclei. The color bar represents the maximum ratio change of an isotopic abundance between the baseline and the changed reaction rate using equation 4. The top panel is system H70Z2M50, the bottom panel is system H50Z1M20. Each panel shows two different reaction types ($p,\gamma$) and ($\alpha, p$) reactions, respectively, from left to right.} 
    \label{fig:heatmap_lc_1}
\end{figure*}

\begin{figure*}[!t]
    \centering
    \includegraphics[width=0.8\textwidth]{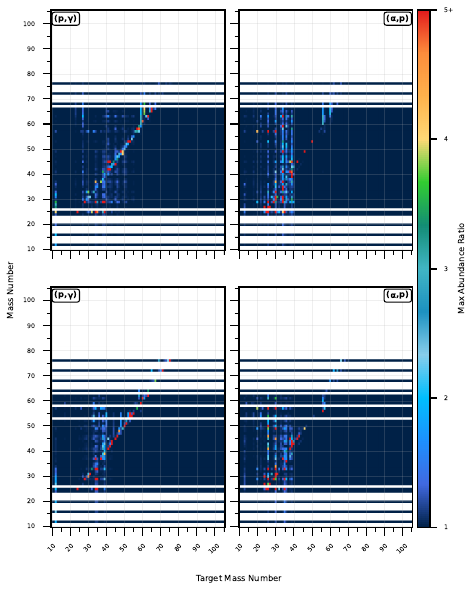}
    \caption{Same as Figure 9. The top panel shows system H30Z2M90, the bottom panel shows system H30Z2M10.} 
    \label{fig:heatmap_lc_2}
\end{figure*}


\section{Discussion} \label{sec:discussion}

In this work, we have integrated three computationally efficient codes: SETTLE to generate semi-analytic temperature and density profiles, NucNet Tools for detailed single-zone nucleosynthetic post-processing, and ONEZONE for obtaining ignition depths and probing reaction-rate sensitivities. The codes were used to perform a sensitivity study of X-ray bursts (XRB) to nuclear reaction rates for a grid of 32 different low-mass X-ray binary systems. By varying over 2,744 proton- and $\alpha$-capture reaction rates within this grid, we pinpoint the key nuclear reaction rates whose uncertainties most significantly shape the burst light curves and nucleosynthetic yields. We identified 49 reactions that significantly affected the burst light curve, and 182 reactions altered the final composition. The outcome aligns with the trends observed in previous studies, where only a reduced number of reactions have a significant impact on the results of the astrophysical models. These important reactions can be categorized based on their role in different parts of the nuclear reaction pathways that drive XRBs (\cite{Meisel_2018}).

The results of Figure~\ref{fig:binaryparms} highlight the dependence of ignition conditions on the system parameters. Higher accretion rate shortens the settling timescale, limiting the extent of stable hydrogen burning and thereby increasing the residual hydrogen fraction at ignition. Simultaneously, lower metallicity depletes the supply of hot CNO catalysts, reducing the efficiency of the hot CNO cycle and further preserving unburned hydrogen. Consequently, with lower metallicity, the thermal instability that triggers the ignition of the XRBs develops deeper into the envelope of the neutron star as compared to metal-rich cases.

We found the most impactful reactions in changing the shape of the light curve are near the bottlenecks of the $\alpha$p- and rp-process pathways. In our sensitivity study, we identified several key reactions in the Ar-K and Ca-Sc mass region that strongly influence the shape of the burst light curve. These include $^{39}\mathrm{K}(\mathrm{p},\gamma)^{40}\mathrm{Ca}$, $^{39}\mathrm{K}(\mathrm{p},\alpha)^{36}\mathrm{Ar}$ (listed as $^{36}\mathrm{Ar}(\alpha,\mathrm{p})^{39}\mathrm{K}$ in Table~\ref{tab:ranked_sensitivity_lc}), $^{38}$Ca($\alpha,\mathrm{p}$)$^{41}$Sc, and $^{39}$Ca($\alpha,\mathrm{p}$)$^{42}$Sc. These reactions facilitate breakout from localized nucleosynthesis cycles formed during the $\alpha$p-process and allow nucleosynthesis to proceed toward heavier nuclei. We also identified four key reactions in the Ni-Ga region that significantly impact the light curve shape: $^{59}\mathrm{Cu}(\mathrm{p},\gamma)^{60}\mathrm{Zn}$, $^{59}\mathrm{Cu}(\mathrm{p},\alpha)^{56}\mathrm{Ni}$ (referenced as $^{56}\mathrm{Ni}(\alpha,\mathrm{p})^{59}\mathrm{Cu}$), $^{63}\mathrm{Ga}(\mathrm{p},\gamma)^{64}\mathrm{Ge}$, and $^{63}\mathrm{Ga}(\mathrm{p},\alpha)^{60}\mathrm{Zn}$ (mentioned as $^{60}\mathrm{Zn}(\alpha,\mathrm{p})^{63}\mathrm{Ga}$). These reactions are deemed as important reactions for the systems with $X_\mathrm{H, ign}$=0.39--0.67, where they influence the burst energetics through their role in localized rp-process cycles.

During the explosive burning in XRBs, localized nucleosynthesis cycles can form. These cycles lead to the recycling of material between nuclei, which can slow down the overall progression of the reaction chains toward heavier elements. In these cycles, we have competition between the ($\mathrm{p},\gamma$) reaction and the ($\mathrm{p},\alpha$) reaction. Such cycles in the $\alpha$p-process path are the Ar-K and the Ca-Sc cycles \cite{1994ApJ...432..326V}. In the Ar-K cycle, the sequence begins with $^{36}\mathrm{Ar}(\mathrm{p},\gamma)^{37}\mathrm{K}(\mathrm{p},\gamma)^{38}\mathrm{Ca}(\beta^+ \nu)^{38}\mathrm{K}(\mathrm{p},\gamma)\\^{39}\mathrm{Ca}(\beta^+ \nu)^{39}\mathrm{K}(\mathrm{p},\alpha)^{36}\mathrm{Ar}$, completing the cycle. The Ca-Sc cycle involves pathways $^{40}\mathrm{Ca}(\mathrm{p},\gamma)^{41}\mathrm{Sc}(\mathrm{p},\gamma)^{42}\mathrm{Ti}(\beta^+,\nu)^{42}\mathrm{Sc}(\mathrm{p},\gamma) ^{43}\mathrm{Ti}\\(\beta^+,\nu)^{43}\mathrm{Sc}(\mathrm{p},\alpha)^{40}\mathrm{Ca}$. The ($\alpha,\mathrm{p}$) reactions play a critical role in leaking from the Ar-K cycle to the Ca-Sc cycle, facilitating the flow of nucleosynthesis towards heavier elements. In our study, we found that the formation of the Ar-K cycle is highly sensitive to the initial hydrogen and helium abundances at the time of burst ignition. When the hydrogen mass fraction ranges from $X_\mathrm{H, ign}$=0.12--0.52, the Ar-K cycle is significantly stronger compared to systems with higher hydrogen content at ignition. The main reactions facilitating leakage from the Ar-K cycle to the Ca-Sc cycle in these models are $^{38}$Ca($\alpha,\mathrm{p}$)$^{41}$Sc, $^{39}$Ca($\alpha,\mathrm{p}$)$^{42}$Sc, $^{39}$K($\mathrm{p},\gamma$)$^{40}\mathrm{Ca}$ reactions. We found a competition between the $^{39}\mathrm{K}(\mathrm{p},\gamma)^{40}\mathrm{Ca}$ and the $^{39}\mathrm{K}(\mathrm{p},\alpha)^{36}\mathrm{Ar}$ in these systems. The down variation of the ($\mathrm{p},\gamma$) reaction has the identical impact as the up variation of the ($\mathrm{p},\alpha$) reaction on the burst light curve. Also, $^{39}\mathrm{Ca}$ can be treated as a bottleneck because of its shell structure, its an equilibrium point where we have competition between ($\mathrm{p},\gamma$) and ($\gamma,\mathrm{p}$) reactions, which makes the proton capture onto the Z+1 nuclei an important reaction to bypass this point. We found other important reactions near these two localized cycles, $^{33}\mathrm{Ar}(\alpha,\mathrm{p})^{36}\mathrm{K}$, $^{37}\mathrm{Ca}(\alpha,\mathrm{p})^{40}\mathrm{Sc}$, $^{39}\mathrm{Ca}(\alpha,\gamma)^{43}\mathrm{Ti}$, and $^{42}\mathrm{Ti}(\alpha,\mathrm{p})^{45}\mathrm{V}$ changes the light curve shape.

The Ni-Cu and Zn-Ga cycles represent important bottlenecks in the rp-process path (\cite{1994ApJ...432..326V}). In the Ni-Cu cycle, competition between $^{59}\mathrm{Cu}(\mathrm{p},\gamma)^{60}\mathrm{Zn}$ and $^{59}\mathrm{Cu}(\mathrm{p},\alpha)^{56}\mathrm{Ni}$ can either allow the reaction flow to continue toward heavier nuclei or redirect it back to $^{56}\mathrm{Ni}$, thereby completing the cycle. Similarly, in the Zn-Ga cycle, the competition between $^{63}\mathrm{Ga}(\mathrm{p},\gamma)^{64}\mathrm{Ge}$ and $^{63}\mathrm{Ga}(\mathrm{p},\alpha)^{60}\mathrm{Zn}$ determines whether the cycle loops back to $^{60}\mathrm{Zn}$ or breaks out toward heavier elements. We find that an increase in the $(\mathrm{p},\alpha)$ reaction rate produces the same effect in the burst's light curve as a decrease, by the same factor, in the competing $(\mathrm{p},\gamma)$ reaction rate. This correlation between competing channels appears to be a general feature across all localized cycles examined in this study.

Our analysis highlights a set of reaction rates at classical waiting point nuclei in the $\alpha$p- and rp-process pathways, whose variations exert substantial influence on the burst light curve. Among them are $(\alpha,\mathrm{p})$ reactions on $^{22}\mathrm{Mg}$, $^{26}\mathrm{Si}$, $^{30}\mathrm{S}$, and $^{34}\mathrm{Ar}$, as well as $(\mathrm{p},\gamma)$ reactions on the Z+1 nuclei, $^{23}\mathrm{Al}$, $^{27}\mathrm{P}$, $^{31}\mathrm{Cl}$, and $^{35}\mathrm{K}$. These reactions occur at waiting point nuclei where the low proton capture $Q$-values and relatively long half-lives establish equilibrium between $(\mathrm{p},\gamma)$ and $(\gamma,\mathrm{p})$ channels. In the intermediate-mass region ($A \approx 20$--$40$), bridging the hot CNO breakout to the rp-process, these nuclei also act as branch points where $(\alpha,\mathrm{p})$ competes with $(\mathrm{p},\gamma)$ reactions. The balance between these two channels determines whether the flow stalls or breaks out toward heavier nuclei. In the rp-process path, the classical waiting points at $^{56}\mathrm{Ni}$ and $^{64}\mathrm{Ge}$ play a critical role. We find that decreasing the rate of $^{56}\mathrm{Ni}(\mathrm{p},\gamma)^{57}\mathrm{Cu}$ and $^{57}\mathrm{Cu}(\mathrm{p},\gamma)^{58}\mathrm{Zn}$ significantly suppresses energy generation in systems with $X_\mathrm{H, ign} = 0.35$--$0.52$. Similarly, $^{65}\mathrm{As}(\mathrm{p},\gamma)^{66}\mathrm{Se}$ affects the burst light curve for higher hydrogen fractions ($X_\mathrm{H, ign} = 0.56\text{--}0.67$), by enabling breakout from the $^{64}$Ge waiting point.

Our sensitivity study identifies two key reactions near the hot CNO region:$^{12}\mathrm{C}(\alpha,\gamma)^{16}\mathrm{O}$ and $^{12}\mathrm{C}(p,\gamma)^{13}\mathrm{N}$, for the He-rich cases. The main breakout reactions, $^{14}\mathrm{O}(\alpha,\mathrm{p})^{17}\mathrm{F}$, $^{15}\mathrm{O}(\alpha,\gamma)^{19}\mathrm{Ne}$, and $^{18}\mathrm{Ne}(\alpha,\mathrm{p})^{21}\mathrm{Na}$, a significant impact was observed only for a few astrophysical conditions.

Our study identifies 182 nuclear reactions whose variations influence the final composition of the XRB ashes across the 32 astrophysical systems we modeled. We classify their effects into two categories: global changes produce broad shifts across multiple mass chains and redistribute the overall nucleosynthetic flow, while local changes are confined to abundances of the immediate parent or daughter nuclei of the modified reaction. The difference between these two types of effects is clearly observed in  Figures~\ref{fig:heatmap_lc_1} and \ref{fig:heatmap_lc_2}. The reactions that lead to local abundance changes appear as high values of abundance sensitivity along the diagonal band in the heatmaps. In contrast, the global changes are evidenced as broad vertical bands that indicate how the abundances of many isotopes are affected by the variation of a single reaction rate, which in most cases correspond to bottleneck reactions in the $\alpha$p and rp‐process paths. 

 A direct comparison of light‐curve and composition sensitivities reveals that only a subset of reactions influences both observables. Most reactions that affect only the light curve leave small localized composition signatures. On the other hand, the majority of reactions that have a high impact only on the composition are ($p,\gamma$) reactions in the medium‐mass ($A>32$) and heavier‐mass ($A>55$) regions that do not alter the burst energetics (Figures~\ref{fig:heatmap_combined}). This separation reflects distinct physical roles: light‐curve–sensitive rates regulate the timescale of energy release, whereas composition‐sensitive rates control the final endpoints of the nucleosynthesis flow. Among the reactions that influence both the light curve and final composition, the most prominent global‐impact cases are the key $\alpha$p‐ and rp‐process bottlenecks. Examples include $^{38}\mathrm{Ca}(\alpha,p)^{41}\mathrm{Sc}$, $^{39}\mathrm{Ca}(\alpha,p)^{42}\mathrm{Sc}$, and $^{39}\mathrm{K}(p,\gamma)^{40}\mathrm{Ca}$ as well as $\alpha$-capture reactions on $^{26}\mathrm{Si}$, $^{30}\mathrm{S}$, and $^{34}\mathrm{Ar}$, and $^{36}\mathrm{Ar}$ all of which produce broad vertical bands in the heatmaps (Figures~\ref{fig:heatmap_lc_1} and \ref{fig:heatmap_lc_2}). In the Ni–Cu and Zn–Ga cycles, the paired competition between ($p,\gamma$) and ($p,\alpha$) reactions on $^{59}\mathrm{Cu}$ and $^{63}\mathrm{Ga}$ produces nearly identical global abundance shifts, mirroring their symmetric effect on the light curve. 

Besides the reactions on the major bottlenecks, the important ones near the breakout and ignition region are $^{12}\mathrm{C}(\alpha,\gamma)^{16}\mathrm{O}$, $^{16}\mathrm{O}(\alpha,\gamma)^{20}\mathrm{Ne}$, and $^{20}\mathrm{Ne}(\alpha,\gamma)^{24}\mathrm{Mg}$. Variations in $^{12}\mathrm{C}(\alpha,\gamma)^{16}\mathrm{O}$ produce localized abundance shifts; downward rate variations diminish local yields, while upward changes increase them across 28 systems. While the up and down variations in $^{16}\mathrm{O}(\alpha,\gamma)^{20}\mathrm{Ne}$ shows the opposite localized effect. The impact of $^{20}\mathrm{Ne}(\alpha,\gamma)^{24}\mathrm{Mg}$ is dependent on the type of variation. Decreasing the rate creates a global shift across He-rich systems, whereas increasing it produces local changes. In the region of the $\alpha$p-process the list of high-impact reactions that leads to local abundance changes is dominated by ($p,\gamma$) and ($\alpha,\gamma$) reactions in the medium‐mass Na–Ca region, including reactions on $^{22-23}\mathrm{Na}$, $^{24}\mathrm{Mg}$, $^{26-27}\mathrm{Al}$, $^{28,30}\mathrm{Si}$, $^{32,34}\mathrm{S}$, $^{34,35}\mathrm{Cl}$, $^{35\text{--}37}\mathrm{Ar}$, $^{37\text{--}39}\mathrm{K}$, and $^{40\text{--}42}\mathrm{Ca}$. In the rp‐process path, most composition sensitivities arise from ($p,\gamma$) reactions in the heavier mass region ($A\geq55$), particularly on nuclei such as $^{55}{\rm Co}$, $^{59}{\rm Cu}$, $^{61}{\rm Ga}$, $^{63}{\rm Ga}$, $^{65}{\rm Ge}$, $^{67}{\rm As}$, $^{69}{\rm Se}$, and $^{71}{\rm Br}$ which produce localized abundance changes. Reactions in this mass region are important in those hydrogen-rich burst models with $X_{H,\mathrm{ign}}\geq0.52$. 
In helium‐rich bursts ($X_{H,\mathrm{ign}}\approx0.01–0.26)$, the reactions that only affect the composition are concentrated on proton captures in the $A\approx32$–54 region, shown in Figure 8. In general, for a reaction with a local impact, an upward variation in the reaction rate depletes the parent isotope in the ashes, while a downward variation increases its yield. Almost all of these $(p, \gamma$) reaction rates are based on the statistical Hauser-Feshbach model \cite{Psaltis_2025}.

One important implication emerges from the subset of reactions that strongly influence the $^{12}$C mass fraction in the ashes, a critical parameter for superburst ignition. The heatmaps of Figures ~\ref{fig:heatmap_lc_1} and ~\ref{fig:heatmap_lc_2} show that a few reactions would cause a significant change in the abundance of $^{12}$C. We have identified systems with low to intermediate hydrogen fractions at ignition ($X_{H,\mathrm{ign}}\approx0.01–0.28)$ where the calculations with modified reaction rates result in  $^{12}$C yields with mass fractions up to 23$\%$. The enhanced yields exceed the 20$\%$ threshold proposed for superburst ignition \citep{2006ApJ...646..429C}. Across these systems, we find 12 reactions with sensitivity to the final $^{12}$C mass fraction, $^{12}$C($p,\gamma$)$^{13}$N, $^{34}$Cl($p,\gamma$)$^{35}$Ar, $^{24}$Mg($\alpha,p$)$^{27}$Al, $^{30}$S($\alpha,p$)$^{33}$Cl, $^{34}$Ar($\alpha,p$)$^{37}$K, $^{34}$Cl($\alpha,p$)$^{37}$Ar, $^{35}$Cl($\alpha,p$)$^{38}$Ar, $^{38}$Ca($\alpha,p$)$^{41}$Sc, $^{26}$Si($\alpha,p$)$^{29}$P, $^{35}$Cl($p,\gamma$)$^{36}$Ar, $^{12}$C($\alpha,\gamma$)$^{16}$O, and $^{24}$Mg($\alpha,\gamma$)$^{28}$Si. These reactions operate primarily in He‐rich bursts, where $\alpha$‐capture chains and key proton captures compete to either preserve or deplete $^{12}$C during late‐time burning. The sensitivity pattern suggests that targeted experimental constraints on these rates would directly refine predictions of $^{12}$C yields and improve assessments of which bursting systems can meet the compositional conditions for superburst ignition. 

 The majority of our sensitivity‐listed reactions have yet to be experimentally constrained in the astrophysical energy range (\cite{RAUSCHER20001}), and reaction network calculations of the rp-process have to rely on uncertain theoretical results, like those from the statistical Hauser-Feshbach model \cite{Psaltis_2025}. The majority of the relevant reactions will remain dependent on theoretical models for the near future, but experimental progress at advanced nuclear physics facilities has opened opportunities for direct and indirect measurements of rp-process reaction rates. Systematic uncertainty quantification and targeted experimental programs guided by sensitivity studies offer a promising path toward reducing these uncertainties to levels required for precision stellar modeling.

\begin{acknowledgments}
We thank Zach Meisel for productive discussions. This work was supported by the US Department of Energy under grant DE-SC0022538  and by the US National Science Foundation under grants PHY-1430152 (JINA-CEE) and OISE-1927130 (IReNA).
\end{acknowledgments}
\clearpage


%



\appendix


\startlongtable
\begin{deluxetable*}{llccccccccccccccccccc}
\tabletypesize{\scriptsize}
\tablecaption{List of impactful reactions due to the change in light curve variation across the 32 binary systems. The table shows the impact factor calculated using the $F_{lc}$ (in units of $10^{17}$ erg $g^{-1}$) criteria for each system corresponding to each reaction variation. N shows the total number of systems for that particular reaction that has a significant impact. We also show the average impact due to the Up and DN variation. The model columns are ordered by hydrogen abundance at ignition (from H-rich to H-poor). The reactions are ordered by their primary element following the periodic table, with multiple reactions for the same element ordered by increasing mass number (from low to high).
\label{tab:ranked_sensitivity_lc}}
\tablehead{
\colhead{\textbf{Rank}} & \colhead{\textbf{Reaction}} &
\multicolumn{16}{c}{\textbf{Models}} &
\colhead{\textbf{N}} & \colhead{$\overline{\text{UP}}$} & \colhead{$\overline{\text{DN}}$} \\
\colhead{} & \colhead{} &
\colhead{\rhead{H70Z1M90}} & \colhead{\rhead{H70Z1M50}} & \colhead{\rhead{H70Z2M90}} & \colhead{\rhead{H70Z2M50}} & \colhead{\rhead{H70Z1M20}} & \colhead{\rhead{H70Z2M20}} & \colhead{\rhead{H50Z1M90}} & \colhead{\rhead{H70Z1M10}} & \colhead{\rhead{H50Z1M50}} & \colhead{\rhead{H50Z2M90}} & \colhead{\rhead{H50Z2M50}} & \colhead{\rhead{H50Z1M20}} & \colhead{\rhead{H70Z2M10}} & \colhead{\rhead{H50Z2M20}} & \colhead{\rhead{H50Z1M10}} & \colhead{\rhead{H70Z1M5}} &
\nocolhead{} & \nocolhead{} & \nocolhead{} \\
\colhead{} & \colhead{} &
\colhead{\rhead{H30Z1M90}} & \colhead{\rhead{H30Z1M50}} & \colhead{\rhead{H30Z2M90}} & \colhead{\rhead{H30Z2M50}} & \colhead{\rhead{H30Z1M20}} & \colhead{\rhead{H50Z2M10}} & \colhead{\rhead{H70Z2M5}} & \colhead{\rhead{H30Z2M20}} & \colhead{\rhead{H50Z1M5}} & \colhead{\rhead{H30Z1M10}} & \colhead{\rhead{H10Z2M90}} & \colhead{\rhead{H10Z1M50}} & \colhead{\rhead{H10Z2M50}} & \colhead{\rhead{H30Z2M10}} & \colhead{\rhead{H50Z2M5}} & \colhead{\rhead{H10Z1M20}} &
\nocolhead{} & \nocolhead{} & \nocolhead{}
}
\startdata
1 & $^{12}\mathrm{C}(\alpha,\gamma)^{16}\mathrm{O}$ &
 -- & -- & -- & -- & -- & -- & -- & -- & -- & -- & -- & -- & -- & -- & -- & 0.6 & 17 & 0.9 & 0.4 \\
 & &
 0.8 & 0.6 & 0.9 & 0.9 & 0.6 & 0.9 & 1.1 & 1.0 & 0.7 & 0.7 & 1.3 & 0.7 & 1.1 & 1.1 & 1.3 & 0.6 & & & \\
\hline
2 & $^{12}\mathrm{C}(p,\gamma)^{13}\mathrm{N}$ &
 -- & -- & -- & -- & -- & -- & -- & -- & -- & -- & -- & -- & -- & -- & -- & -- & 14 & 0.6 & 0.4 \\
 & &
 0.6 & -- & 0.6 & 0.6 & -- & 0.6 & 0.7 & 0.7 & 0.5 & 0.4 & 0.9 & 0.5 & 0.7 & 0.8 & 0.9 & 0.4 & & & \\
\hline
3 & $^{13}\mathrm{N}(\alpha,p)^{16}\mathrm{O}$ &
 -- & -- & -- & -- & -- & -- & -- & -- & -- & -- & -- & -- & -- & -- & -- & -- & 4 & 0.0 & 0.3 \\
 & &
 -- & -- & -- & -- & -- & -- & -- & -- & -- & -- & 0.3 & -- & 0.3 & 0.3 & 0.3 & -- & & & \\
\hline
4 & $^{14}\mathrm{O}(\alpha,p)^{17}\mathrm{F}$ &
 -- & -- & -- & -- & -- & -- & -- & -- & -- & -- & -- & -- & -- & -- & -- & -- & 1 & 0.4 & 0.0 \\
 & &
 -- & -- & -- & -- & -- & -- & 0.4 & -- & -- & -- & -- & -- & -- & -- & -- & -- & & & \\
\enddata
\end{deluxetable*}

\startlongtable
\begin{deluxetable*}{llccccccccccccccccccc}
\tabletypesize{\scriptsize}
\tablecaption{List of impactful reactions due to the change in final abundance across the 32 binary systems. 
\label{tab:ranked_sensitivity_ash}
}
\tablehead{
\colhead{\textbf{Count}} & \colhead{\textbf{Reaction}} &
\multicolumn{16}{c}{\textbf{Models}} &
\colhead{\textbf{N}} & \colhead{$\overline{\text{UP}}$} & \colhead{$\overline{\text{DN}}$} \\
\colhead{} & \colhead{} &
\colhead{\rhead{H70Z1M90}} & \colhead{\rhead{H70Z1M50}} & \colhead{\rhead{H70Z2M90}} & \colhead{\rhead{H70Z2M50}} & \colhead{\rhead{H70Z1M20}} & \colhead{\rhead{H70Z2M20}} & \colhead{\rhead{H50Z1M90}} & \colhead{\rhead{H70Z1M10}} & \colhead{\rhead{H50Z1M50}} & \colhead{\rhead{H50Z2M90}} & \colhead{\rhead{H50Z2M50}} & \colhead{\rhead{H50Z1M20}} & \colhead{\rhead{H70Z2M10}} & \colhead{\rhead{H50Z2M20}} & \colhead{\rhead{H50Z1M10}} & \colhead{\rhead{H70Z1M5}} &
\nocolhead{} & \nocolhead{} & \nocolhead{} \\
\colhead{} & \colhead{} &
\colhead{\rhead{H30Z1M90}} & \colhead{\rhead{H30Z1M50}} & \colhead{\rhead{H30Z2M90}} & \colhead{\rhead{H30Z2M50}} & \colhead{\rhead{H30Z1M20}} & \colhead{\rhead{H50Z2M10}} & \colhead{\rhead{H70Z2M5}} & \colhead{\rhead{H30Z2M20}} & \colhead{\rhead{H50Z1M5}} & \colhead{\rhead{H30Z1M10}} & \colhead{\rhead{H10Z2M90}} & \colhead{\rhead{H10Z1M50}} & \colhead{\rhead{H10Z2M50}} & \colhead{\rhead{H30Z2M10}} & \colhead{\rhead{H50Z2M5}} & \colhead{\rhead{H10Z1M20}} &
\nocolhead{} & \nocolhead{} & \nocolhead{}
}
\startdata
1 & $^{12}\mathrm{C}(p,\gamma)^{13}\mathrm{N}$ &
 9 & 11 & -- & 10 & 13 & 11 & 9 & 7 & 7 & 9 & 7 & -- & 6 & -- & -- & -- & 11 & 9 & 6 \\
 & &
 -- & -- & -- & -- & -- & -- & -- & -- & -- & -- & -- & -- & -- & -- & -- & -- & & & \\
 & &
 16 & 16 & -- & 16 & 16 & 16 & 20 & 20 & 20 & 20 & 24 & -- & 24 & -- & -- & -- & & & \\
 & &
 -- & -- & -- & -- & -- & -- & -- & -- & -- & -- & -- & -- & -- & -- & -- & -- & & & \\
\tablebreak
2 & $^{12}\mathrm{C}(\alpha,\gamma)^{16}\mathrm{O}$ &
 -- & -- & -- & -- & 13 & 36 & 46 & 42 & 41 & 42 & 35 & 32 & 23 & 39 & 55 & 66 & 28 & 17 & 50 \\
 & &
 71 & 71 & 69 & 69 & 67 & 66 & 69 & 67 & 60 & 59 & 62 & 59 & 65 & 67 & 66 & 76 & & & \\
 & &
 -- & -- & -- & -- & 16 & 16 & 16 & 16 & 16 & 16 & 16 & 16 & 16 & 16 & 16 & 16 & & & \\
 & &
 16 & 16 & 16 & 16 & 16 & 16 & 16 & 16 & 16 & 16 & 16 & 16 & 16 & 16 & 16 & 16 & & & \\
\enddata
\tablecomments{%
The table shows the impact factor using the $F_{ash}$ criteria for each system and the highest mass value changed due to the criteria. For each reaction row, the top two rows show the sensitivity factor calculated using the criteria, and the bottom two rows show the highest mass value change across each system. The values with a very large factor are due to the vanishing of one isotopic abundance, which is written as of the order of 10. $N$ is the number of systems where that reaction is significant. We also report the average impact for UP and DN variations.
}
\end{deluxetable*}

\newpage

\bibliography{xrb}
\bibliographystyle{aasjournal}



\end{document}